\documentclass{jfm}
\usepackage{amsmath}
\usepackage{siunitx}
\usepackage{xcolor}
\usepackage{newtxtext}
\usepackage{newtxmath}
\usepackage{natbib}
\newcommand{\RomanNumeralCaps}[1]
\linenumbers

\graphicspath{{./figures_final/}}

\usepackage[normalem]{ulem}

\title{Lagrangian diffusion properties of a free shear turbulent jet}
\shorttitle{Lagrangian diffusion properties of a free shear turbulent jet}

\author{Bianca Viggiano\aff{1} \corresp{\email{viggiano@pdx.edu}}, Thomas Basset\aff{2}, Stephen Solovitz\aff{3}, Thomas Barois\aff{4}, Mathieu Gibert\aff{5}, Nicolas Mordant\aff{6}, Laurent Chevillard\aff{2}, Romain Volk\aff{2}, Micka\"el Bourgoin\aff{2} and Ra\'ul Bayo\'an Cal\aff{1}}
\affiliation{\aff{1}Department of Mechanical and Materials Engineering, Portland State University, Portland, OR 97201, USA
\aff{2}Univ Lyon, ENS de Lyon, Univ Claude Bernard, CNRS, Laboratoire de Physique, 69007 Lyon, France
\aff{3}School of Engineering and Computer Science, Washington State University Vancouver, Vancouver, WA 98686, USA
\aff{4}University of Bordeaux, CNRS, LOMA, 33400 Talence, France
\aff{5}Univ. Grenoble Alpes, CNRS, Grenoble INP, Institut Néel, 38000 Grenoble, France
\aff{6}Univ. Grenoble Alpes, CNRS, Grenoble INP, LEGI, 38000 Grenoble, France}
\shortauthor{B. Viggiano and others}

\begin{document}

\maketitle

\begin{abstract}
A Lagrangian experimental study of an axisymmetric turbulent water jet is performed to investigate the highly anisotropic and inhomogeneous flow field. The measurements were conducted within a Lagrangian exploration module, an icosahedron apparatus, to facilitate optical access of three cameras. The stereoscopic particle tracking velocimetry results in three component tracks of position, velocity and acceleration of the tracer particles within the vertically-oriented jet with a Taylor-based Reynolds number $\Rey_\lambda \simeq 230$. Analysis is performed at seven locations from 15 diameters up to 45 diameters downstream. Eulerian analysis is first carried out to obtain critical parameters of the jet and relevant scales, namely the Kolmogorov and large turnover (integral) scales as well as the energy dissipation rate. Lagrangian statistical analysis is then performed on velocity components stationarised following methods inspired by Batchelor (\textit{J. Fluid Mech.}, vol. 3, 1957, pp. 67-80) which aim to extend stationary Lagrangian theory of turbulent diffusion by Taylor to the case of self-similar flows. The evolution of typical Lagrangian scaling parameters as a function of the developing jet is explored and results show validation of the proposed stationarisation. The universal scaling constant $C_0$ (for the Lagrangian second-order structure function), as well as Eulerian and Lagrangian integral time scales are discussed in this context. $C_0$ is found to converge to a constant value (of the order of $C_0 = 3$) within 30 diameters downstream of the nozzle. Finally, the existence of finite particle size effects are investigated through consideration of acceleration dependent quantities.
\end{abstract}

\section{Introduction}
The dispersion of particles from a point source in turbulent free jet flows plays an important role in many industrial and natural systems, including for instance sprays, flames, volcanic flows, and emission of pollutants at industrial chimneys. Depending on the particle characteristics (size, density with respect to the carrier fluid, volume fraction, etc.), its dynamics will follow that of the fluid (particles will then be considered as tracers) or it may be affected by inertial effects, finite size effects and couplings between the phases in highly seeded particle-laden flows \citep{berk2020transport}.

In the simplest situations, where particles can be considered as tracers (which is the framework of the present study), the turbulent diffusion process can be related to simple Lagrangian statistical properties of the carrier flow. While this connection has been extensively investigated for the case of homogeneous isotropic turbulence, in the spirit of Taylor's turbulent diffusion theory \citep{taylor1922diffusion}, the case of inhomogeneous flows remains largely unexplored, in spite of an extension of Taylor's theory proposed by \citet{batchelor1957diffusion}. A summary of Taylor's diffusion and Batchelor's extension to self-similar flows are presented herein as incentive for the characterisation of several basic Lagrangian statistics in free shear flows and in turn, motivation of the present study.

\subsection{Taylor's theory of turbulent diffusion}
The importance of the Lagrangian approach to model turbulent dispersion was first evidenced by the early work of \citet{taylor1922diffusion}. Taylor's theory connects the mean square displacement (MSD) $\sigma^2(\tau)$ of particles spreading from a point source in stationary homogeneous isotropic turbulence to the Lagrangian two-point correlation function $R^L_{uu}(\tau) = \langle u(t+\tau)u(t) \rangle$, where the average $\langle \cdot \rangle$ is taken over an ensemble of particle trajectories. Here, $u(t)$ represents the velocity of individual particles along their trajectory (note that for simplicity only one velocity component is considered) and $\tau$ is the time lag. This result is often called the Taylor theorem and expressed as:
\begin{equation}
\frac{\mathrm{d}^2\sigma^2}{\mathrm{d}\tau^2}(\tau) = 2 R^L_{uu}(\tau).
\label{eq:taylor}
\end{equation}
Taylor's theory is of utmost practical importance, as it reduces the prediction of the spreading of tracer particles (and therefore of any passive substance spread by turbulence with negligible molecular diffusivity) to the knowledge of the Lagrangian two-point correlation function $R^L_{uu}(\tau)$ at all times. Note that the correlation function $R^L_{uu}$ can be equivalently replaced by the Lagrangian second-order structure function $S^L_2(\tau) = \langle [u(t+\tau)-u(t)]^2 \rangle = 2 (R^L_{uu}(0) - R^L_{uu}(\tau))$, which is a common statistical tool used to characterise the multi-scale dynamics of turbulence. The correlation at $\tau = 0$, $R^L_{uu}(0)$, is the mean square of the velocity fluctuations $\sigma_u^2$.

Interestingly, the asymptotic regimes of the short and long time scales of turbulent diffusion do not depend on the details of the dynamics of turbulence. In the limit of very short times, the spreading follows trends of the trivial (purely kinematic) ballistic regime where $\sigma^2(\tau) \simeq \sigma_u^2 \tau^2$. This can be retrieved from a simple one term Taylor expansion of the particle displacement itself, or equivalently by applying equation~\eqref{eq:taylor} and considering the limit at vanishing times for the Lagrangian correlation function, $R^L_{uu}(\tau) \simeq \sigma_u^2$ for small times. In the limit of very long time scales, equation~\eqref{eq:taylor} from Taylor's theory predicts that due to the finite Lagrangian correlation time of turbulence ($T_L = \sigma_u^{-2} \int_0^\infty R^L_{uu}(\tau) \:\mathrm{d}\tau$) the long term turbulent diffusion process behaves as simple diffusion (where the MSD grows linearly with time, $\sigma^2 \propto 2 K_T \tau$, for long times) with a turbulent diffusivity $K_T = \sigma_u^2 T_L$.

Detail of the diffusion process at intermediate time scales requires a deeper knowledge of the specific time dependence of $R^L_{uu}(\tau)$ at all times, particularly in the inertial range of scales of turbulence. Such dependency can be inferred empirically from a Lagrangian statistical description \emph{\`a la} Kolmogorov \citep{toschi2009lagrangian}, which predicts that for homogeneous isotropic turbulence within the inertial range of time scales, $\tau_\eta \ll \tau \ll T_L$, $S^L_2(\tau) = C_0 \varepsilon \tau$, with $\varepsilon$ the turbulent energy dissipation rate and $\tau_\eta = (\nu/\varepsilon)^{1/2}$ the turbulent dissipation scale. The universal constant $C_0$ plays a similar role in the Lagrangian framework to the Kolmogorov constant in the Eulerian framework. As a consequence, a detailed description of the turbulent diffusion process, including the inertial scale behaviour, relies on the knowledge of $S^L_2(\tau)$ (or equivalently of $R^L_{uu}(\tau)$) at all time scales and specifically on the knowledge of $C_0$ at inertial scales. Thereafter, stochastic models can be built giving reasonable Lagrangian dynamics descriptions at all time scales \citep{sawford1991reynolds, viggiano2020modelling}.

The empirical determination of the constant $C_0$ is therefore critical in describing the turbulent diffusion process and to accurately model the particle dispersion occurring in industrial applications and natural circumstances. Such a determination requires accessing accurate inertial range Lagrangian statistics and has received attention in the past two decades in several experimental and numerical studies \citep{sawford1991reynolds, mordant2001measurement, yeung2002lagrangian, ouellette2006small, toschi2009lagrangian} as well as some field measurements in the ocean \citep{lien1998lagrangian}. This leads to a range of $C_0$ estimates ranging from 2 to 7 (c.f. \citet{lien2002kolmogorov} and \citet{toschi2009lagrangian} for a complete comparison of theoretical, simulated and experimental results). The variability of reported values in literature have been in part attributed to the relatively strong dependence of this constant on Reynolds number \citep{sawford1991reynolds, ouellette2006small} and on the existence of large scale anisotropy and inhomogeneity \citep{ouellette2006small}.
 
\subsection{Batchelor's extension of theory of turbulent diffusion}
In spite of this variability of the tabulated values for $C_0$, the connection between turbulent diffusion and Lagrangian statistics in homogeneous isotropic and stationary turbulence is now well circumscribed. The situation is more complex when it comes to inhomogeneous and anisotropic flows. One strong hypothesis of Taylor's turbulent diffusion theory relies on the statistical Lagrangian stationarity of the particle dynamics, which requires not only a global temporal stationarity of the flow, but also a statistical Eulerian homogeneity: a particle travelling across an inhomogeneous field will indeed experience non-stationary temporal dynamics along its trajectory. Besides, in inhomogeneous flows any Lagrangian statistics will depend on the initial position of the particle (used to label trajectories). For sake of keeping formulas compact, explicit reference to initial position will be omitted when exploring inhomogeneous Lagrangian statistics, but the reader should remember this dependence.

One such inhomogeneous flow field is the turbulent free round jet. Although limited Lagrangian experimental campaigns has been carried out \citep{holzner2008lagrangian,wolf2012investigations,kim2017characterisation,gervais2007acoustic}, this type of flow has received much attention in Eulerian studies as one of its most striking properties is that turbulence is self-preserving \citep{corrsin1943investigation, hinze1949transfer, hussein1994velocity, weisgraber1998turbulent}. More specifically, as the jet develops downstream of the nozzle, the turbulence properties (length, time and velocity scales) evolve in such a way that the Reynolds number remains constant at all downstream positions. Note that such self-similarity generally applies only at sufficiently large downstream positions, typically $z \gtrsim 20 D$, with $D$ the nozzle diameter \citep{pope2000turbulent}. As a result of this axial Eulerian inhomogeneity, Lagrangian dynamics is non-stationary and dependent on the initial position of considered trajectories.

In 1957, Batchelor proposed an extension of Taylor's stationary diffusion theory to the case of turbulent jets in a Lagrangian framework, exploiting the Eulerian self-similarity property of these flows \citep{batchelor1957diffusion}. The approach by Batchelor uses the Eulerian self-similarity to define a \textit{compensated} time $\tilde{\tau}$ and a \textit{compensated} Lagrangian velocity $\tilde{u}(\tilde{\tau})$ which exhibits statistically stationary Lagrangian dynamics. It can be noted that the Lagrangian stationarisation idea introduced by Batchelor is not limited to the case of the jet, but can also be applied to other self-preserving flows such as wakes, mixing layers and possibly other types of shear flows \citep{batchelor1957diffusion, cermak1963lagrangian}.

The idea of this stationarisation is to compensate the effect of Eulerian inhomogeneity on the Lagrangian variables to retrieve a Lagrangian dynamics which becomes independent of the initial position and statistically stationary and in turn, to generalise results originally established for stationary situations (such as Taylor's theory of turbulent diffusion). Based on the Eulerian self-similarity properties, Batchelor considers the case of the dispersion of particles released at the origin of a turbulent jet, whose Lagrangian dynamics is stationarised by considering the just mentioned compensated variables. Explicitly, through consideration of the velocity at the position $\boldsymbol{x}(\tau)$ reached by the particle at a given time $\tau$ since it has been released (at $\tau = 0$ and $\boldsymbol{x} = 0$) as well as the time scales of the flow properties at this position $\boldsymbol{x}(\tau)$:
\begin{equation}
\tilde{u}(\tau) = \frac{u(\tau) - \overline{u}^e(\boldsymbol{x}(\tau))}{\sigma_u(\boldsymbol{x}(\tau))} \quad \textrm{and} \quad \tilde{\tau} = \frac{\tau}{T_E(\boldsymbol{x}(\tau))},
\label{eq:ucomp}
\end{equation}
where $\overline{u}^e(\boldsymbol{x}(\tau))$ represents the local (Eulerian) average velocity at the position $\boldsymbol{x}$ of the particle at time $\tau$ and $T_E(\boldsymbol{x}(\tau))$ the local Eulerian time scale (only one velocity component is considered). Similarly, $\sigma_u(\boldsymbol{x}(\tau))$ is the local (Eulerian) standard deviation of the velocity at the position $\boldsymbol{x}$ of the particle at time $\tau$. The temporal transformation simply rescales the time in order to account for the evolution of the Eulerian background properties as the particle moves downstream in the jet. The transformation of the velocity intends to stationarise the effective dynamics by: (i)~subtracting the local average velocity, so that the average of $\tilde{u}$ is zero, and (ii)~the denominator $\sigma_u(\boldsymbol{x}(\tau))$ is chosen as a general compensation for the decay of the turbulent fluctuations of the background Eulerian field as the particles moves downstream. Note that the transformations, as they were presented by \citet{batchelor1957diffusion}, directly considered the Eulerian power-law dependencies (in space) of $\overline{u}^e$, $\sigma_u$ and $T_E$ in the self-similar region of the jet near its centerline. The transformations as written in equations~\eqref{eq:ucomp} are therefore more general, although Batchelor's transformations are eventually equivalent if such power-law dependencies are assumed. The more general expression considered here allows one to explore the relevance of the stationarisation procedure not only in the centerline of the jet (as done by Batchelor) but to also probe away of the centerline.

As a result of the stationarisation procedure, compensated Lagrangian statistics are expected to no longer depend on the initial position and to exhibit similar properties (time scales, correlations, etc.) at any position in the jet and hence at any time along particle trajectories. Batchelor then demonstrates that Taylor's theory can be extended to the stationarised dynamics by connecting the mean square displacement of the particles to $R^L_{\tilde{u}\tilde{u}}(\tilde{\tau})$, the Lagrangian correlation function of $\tilde{u}(\tilde{\tau})$.

Three important aspects arise regarding Batchelor's diffusion theory: (i)~it extends the Eulerian self-similarity to the Lagrangian framework, with this respect it is often referred to as Lagrangian self-similarity hypothesis \citep{cermak1963lagrangian}, (ii)~it connects the turbulent diffusion process of particles in jets to the Lagrangian correlation function (or equivalently to the second-order structure function) of the stationarised velocity statistics and (iii)~it proposes a systematic method of analysing the non-stationary data of the jet.

\subsection{Outline of the article}
To the knowledge of the authors, only indirect evidence of the validity concerning Batchelor's self-similarity hypothesis in turbulent free jets exists in the literature, largely based on measurements of the mean square displacements of particles \citep{kennedy1998particle}. Direct Lagrangian measurements which show the stationarity of the compensated velocity correlations are still lacking, as well as the full characterisation of the inertial scale Lagrangian dynamics in jets. Lagrangian correlation functions in free shear jets have been reported in experiments by \citet{gervais2007acoustic}, using acoustic Lagrangian velocimetry \citep{mordant2001measurement}, although the question of the Lagrangian self-similarity has not been addressed, neither has the detailed characterisation of the inertial range dynamics, the estimation of the related fundamental constants such as $C_0$, and the relevance of simple Lagrangian stochastic models derived for homogeneous isotropic conditions \citep{sawford1991reynolds}.

The aim of the present article is to address these unanswered questions through examination of particle trajectories within a free jet. Three component trajectories of a turbulent water jet ($\Rey_\lambda \simeq 230$) are captured, with a measurement volume containing up to 45 diameters downstream of the jet exit. Experimental methods provide sufficient temporal details to analyse particle trajectories as well as adequate spatial resolution and interrogation volume size to facilitate the application of basic Eulerian analysis. In section~\ref{sec:exp} the experimental setup and methods are presented including the implementation of the Lagrangian particle tracking and the stationarisation procedure that will be applied, inspired by Batchelor's Lagrangian self-similarity hypothesis. Section~\ref{sec:eul} is dedicated to basic Eulerian statistics, which are not the main topic of this study but nevertheless allow the characterisation of key turbulence properties (energy dissipation rate, Eulerian scales, Reynolds number, etc.) and their self-similar behaviour. Section~\ref{sec:lag} includes results on the Lagrangian dynamics. In the context of the previously discussed turbulent diffusion, emphasis is placed on second-order Lagrangian statistics (velocity two-point correlation and structure functions), for which the Lagrangian self-similarity compensation is tested and the estimate of the constant $C_0$ is given. The connections between Eulerian and Lagrangian scales are also considered in the framework of classical stochastic modelling. Section~\ref{sec:acc} extends the discussion of Lagrangian statistics to second-derivative dynamics where comparisons between key acceleration quantities and the scaling constant, $C_0$, are presented. Finally, main conclusions are summarised in section~\ref{sec:conc}.

\section{Experimental methods}\label{sec:exp}
\subsection{Hydraulic setup}
\begin{figure}
\hspace{0.03\textwidth}(\textit{a}) \hspace{0.32\textwidth}(\textit{b})
\begin{center}
\includegraphics[height=0.4\textwidth]{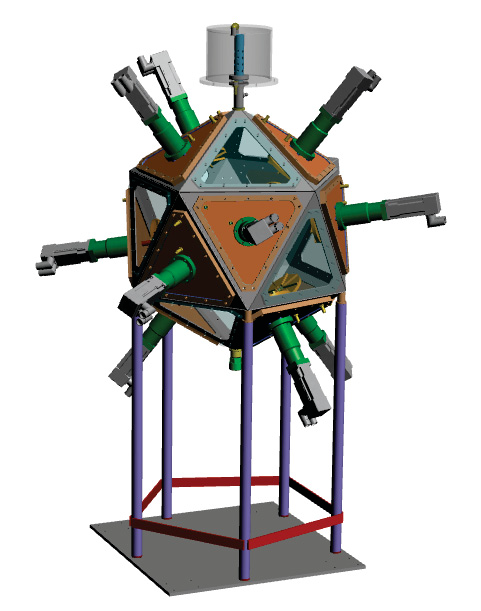}
\hspace{0.5cm}
\includegraphics[height=0.4\textwidth]{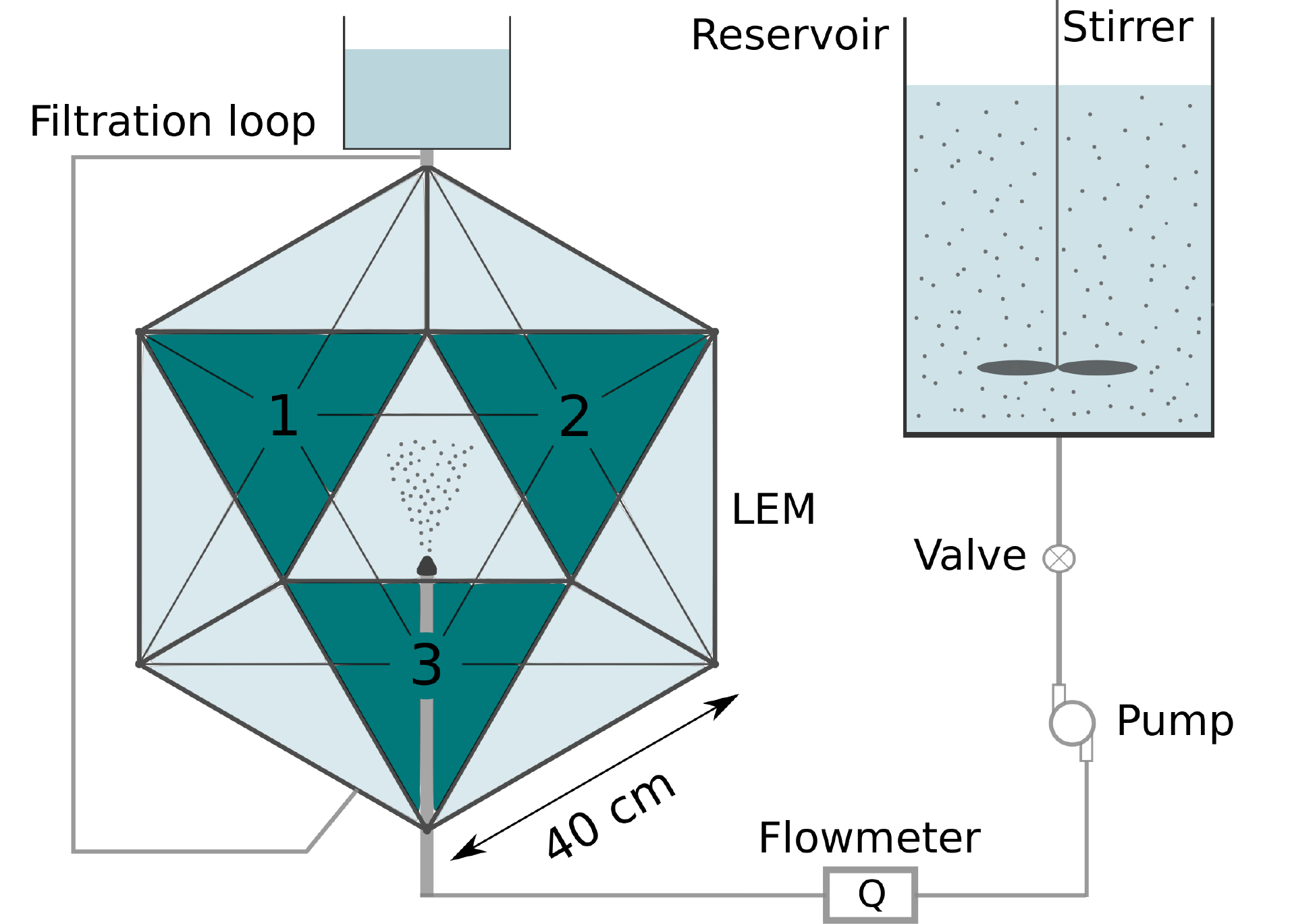} 
\end{center}
\caption{(a)~Three-dimensional CAD rendering of the Lagrangian Exploration Module. (b)~Schematic of the hydraulic setup. Cameras 1, 2 and 3 are oriented orthogonal to the green faces labelled accordingly as 1, 2 and 3.\label{fig:lem_draw_schem}}
\end{figure}
Experiments were performed in the Lagrangian exploration module (LEM) \citep{zimmermann2010lagrangian} at the \'Ecole Normale Sup\'erieure de Lyon. A vertically-oriented jet of water is injected into the LEM, a convex regular icosahedral (twenty-faced polyhedron) tank full of water, as seen in figure~\ref{fig:lem_draw_schem}(a). The LEM is originally designed to generate homogeneous isotropic turbulence when the twelve propellers on twelve of its faces are activated, however, for this experiment, the LEM is only used as a tank as the optical access makes it an ideal apparatus for three-dimensional particle tracking of a jet.

A schematic of the hydraulic setup is included in figure~\ref{fig:lem_draw_schem}(b). The vertical jet, injected with a pump connected to a reservoir, is ejected from a round nozzle with a diameter $D = \SI{4}{mm}$. At the nozzle exit, the flow rate is kept steady at $Q \simeq \SI{e-4}{m^3/s}$, generating an exit velocity $U_J \simeq \SI{7}{m/s}$, and in turn, a Reynolds number based on the diameter $\Rey_D = U_JD/\nu \simeq \SI{2.8e4}{}$ with $\nu$ the water kinematic viscosity. Experiments are performed at ambient temperature. By moving the vertical position of the nozzle, two locations are considered in order to study near-field (NF) and far-field (FF) dynamics, with interrogation volumes spanning from $\SI{0}{mm} \leq z \leq \SI{120}{mm}$ ($0 \leq z/D \leq 30$) and $\SI{80}{mm} \leq z \leq \SI{200}{mm}$ ($20 \leq z/D \leq 50$), respectively (the $z$-axis is the jet axis with $z = 0$ the nozzle exit position). For both regions, the jet is sufficiently far from the walls of the tank to discount momentum effects from the LEM onto the jet \citep{hussein1994velocity}, and thus a free jet is observed.

The particles, seeding the jet during injection, are neutrally buoyant spherical polystyrene tracers with a density $\rho_p = \SI{1060}{kg/m^3}$ and a diameter $d_p = \SI{250}{\micro\meter}$. The reservoir is seeded with a mass loading of $0.1\%$ (reasonable seeding to observe a few hundred particles per image) and an external stirrer maintains homogeneity of the particles. The quiescent water inside the LEM is not seeded, therefore tracked particles are only those injected into the measurement volume by the jet (although some tracers are always remaining in the tank). The inlet valve is open some seconds before the recording, in such a way that the jet is stationary but minimal particle recirculation occurs. The ratio of the particles diameter $d_p$ to the Taylor microscale $\lambda$ is always smaller than 1 and ranges from 0.3 (in the far-field) to 0.8 (in the near-field). The particles are not expected to deviate from tracer behaviour for velocity statistics within the inertial range \citep{mordant2004experimental1}. The ratio $d_p$ to the Kolmogorov length scale $\eta$ remains, however, larger than 1 and ranges from 9 (in the far-field) to 25 (in the near-field). Finite size effects are therefore expected to influence small scale Lagrangian dynamics and in particular acceleration statistics \citep{qureshi2007turbulent}, as further investigated in section~\ref{sec:acc}.

\subsection{Optical setup}
\begin{figure}
\hspace{0.02\textwidth}(\textit{a}) \hspace{0.4\textwidth}(\textit{b})
\begin{center}
\begin{minipage}[c]{0.48\linewidth}
\includegraphics[scale=0.5]{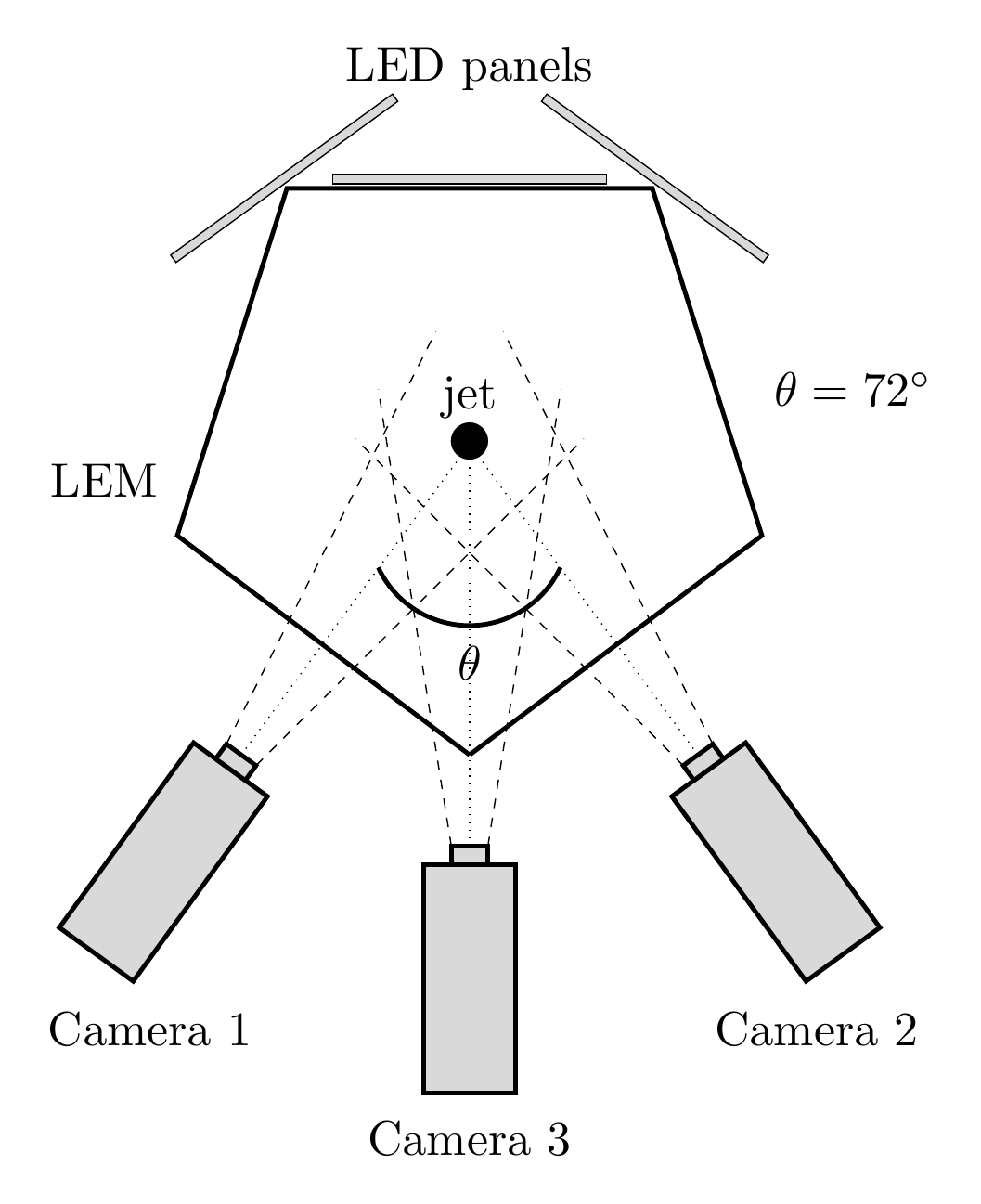}
\end{minipage}
\begin{minipage}[c]{0.48\linewidth}
\hspace*{-1cm}
\includegraphics[scale=0.5]{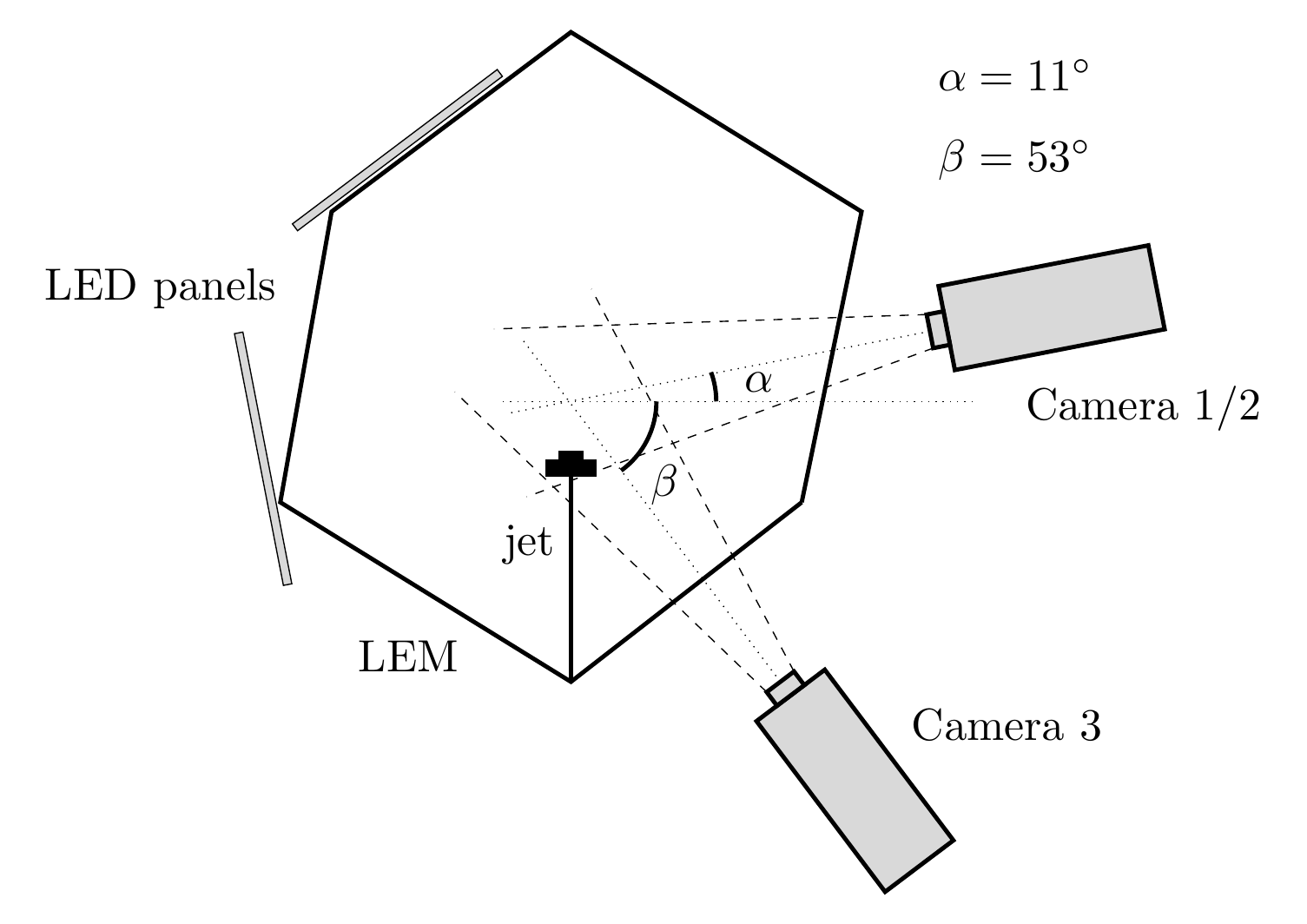}
\end{minipage} 
\end{center}
\caption{Schematic of the optical setup: (a)~top view and (b)~profile view.\label{fig:optical_setup}}
\end{figure}
\begin{figure}
\centerline{\includegraphics[width=0.6\textwidth]{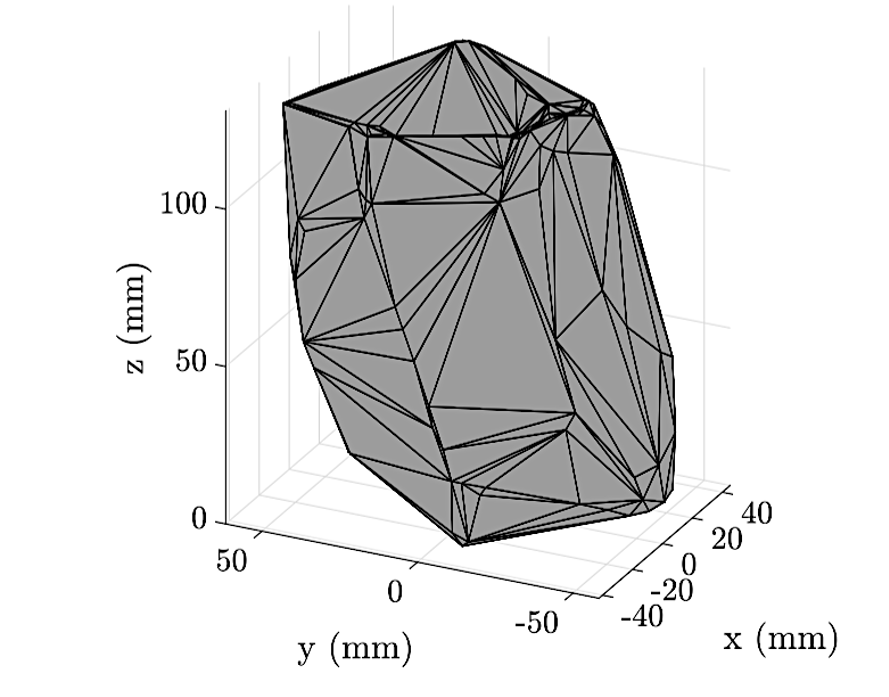}}
\caption{Measurement volume captured by the three camera setup for the near-field measurements (same measurement volume for the far-field measurements).\label{fig:convhull}}
\end{figure}
Three high speed cameras (Phantom V12, Vision Research) mounted with $\SI{100}{mm}$ macro lenses (ZEISS Milvus) are used to track the particles. The optical configuration is shown in figure~\ref{fig:optical_setup}. The angles are related to the geometry of an icosahedron. The interrogation volume is illuminated in a back light configuration with three $\SI{30}{cm}$ square LED panels oriented opposite the three cameras. The spatial resolution of each camera is $1280 \times 800$ pixels, creating a measurement volume of around $80 \times 100 \times \SI{130}{mm^3}$, as seen in figure~\ref{fig:convhull}, hence one pixel corresponds to roughly $\SI{0.1}{mm}$. The three cameras are synced via TTL triggering at a frequency of $\SI{6}{kHz}$ for 8000 snapshots, resulting in a total record of nearly $\SI{1.3}{s}$ per run. For each nozzle position (NF and FF), a total of 50 runs are performed to ensure statistical convergence.

\subsection{Particle Tracking Velocimetry (PTV)}
\subsubsection{Particle detection}
\begin{figure}
\centerline{\includegraphics[width=\textwidth]{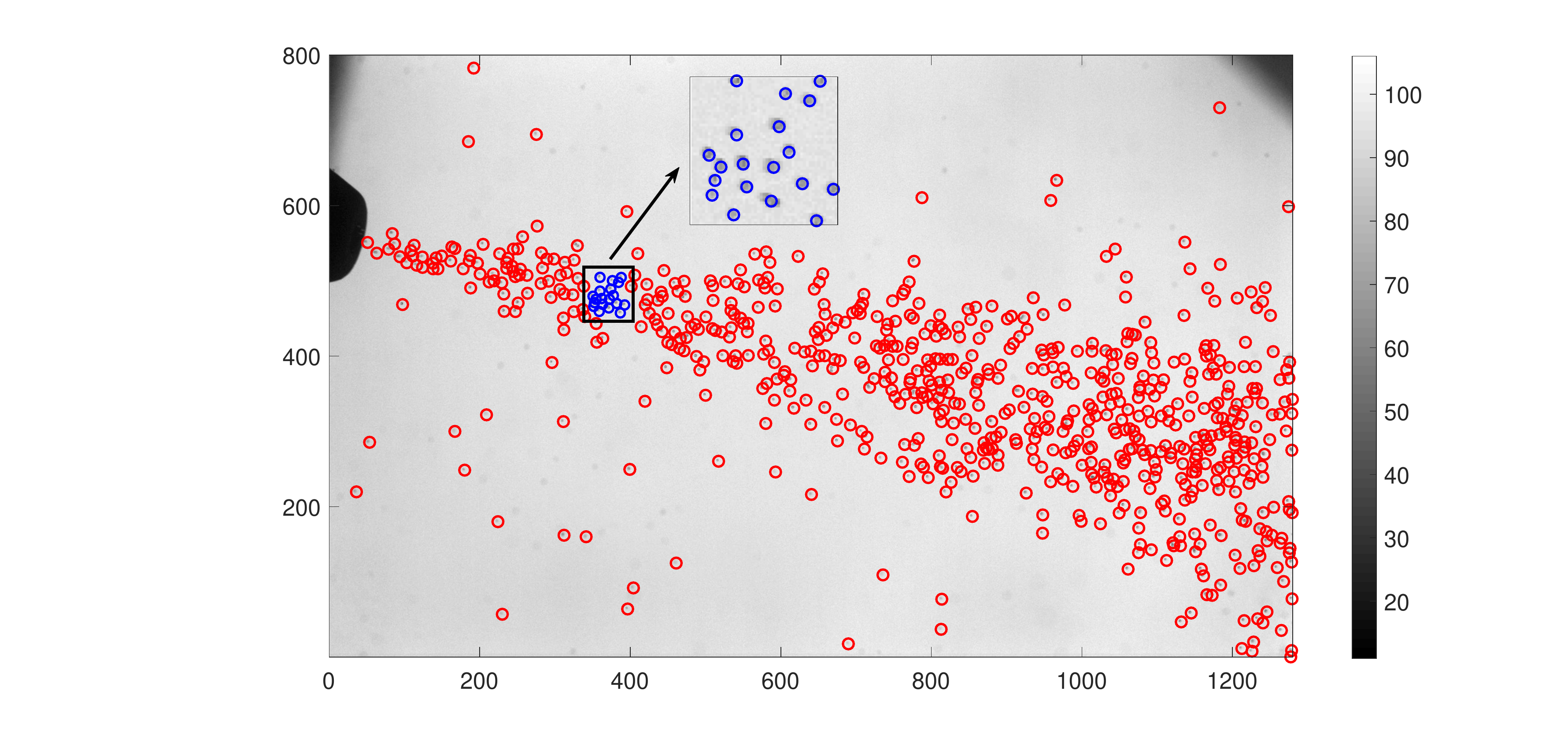}}
\caption{Detection of 705 particles on camera 2 in the near-field configuration (nozzle in the top left-hand corner). Inset: zoom on the boxed zone.\label{fig:im_proc}}
\end{figure}
To create particle trajectories through PTV, two-dimensional images are first analysed to measure the positions of the centres of the particles. The particle detection procedure used in this study is an \textit{ad hoc} process which uses classical methods of image analysis: non-uniform illumination correction, morphological operations (opening), thresholding, binarisation and centroid detection. An example of the camera image with detected particles is presented in figure~\ref{fig:im_proc}.

\subsubsection{Stereoscopic reconstruction}
After the particle centres for all images and all cameras have been determined, the actual three-dimensional positions of the particles can be reconstructed, knowing that each camera image is a two-dimensional projection of the measurement volume. More typically, methods based on optical models are used to achieve real particle positions, but for this study a geometric method developed by \citet{machicoane2019simplified} is used due to its increased precision and ease of implementation. This method is based on an initial polynomial calibration, where each position on a camera image corresponds to a line in real space (a line of possible positions in three-dimensional space). The rays for each detected centre in the two-dimensional images are computed based on the calibration then those rays are matched in space for all three camera locations to create a volume of particles in real space. The matching algorithm employed was recently developed by \citet{bourgoin2020using}. To create the largest convex hull possible which is dictated by the orientation of the cameras, matching of particle position based on the intercept of only two rays (i.e. two cameras of the three total) is accepted. The possibility of overlapping of particles in one dimension, two matches per ray, is also admitted in this algorithm. However this allows the inclusion of non-existent ghost particles. Fortunately, these ghost particles do not form persistent trajectories and therefore they are removed when the trajectories are formed in the next step. The tolerance to allow a match is $\SI{50}{\micro\meter}$ (calibration accuracy around $\SI{1}{\micro\meter}$).

\subsubsection{Tracking}
\begin{figure}
\centerline{\includegraphics[width=0.6\textwidth]{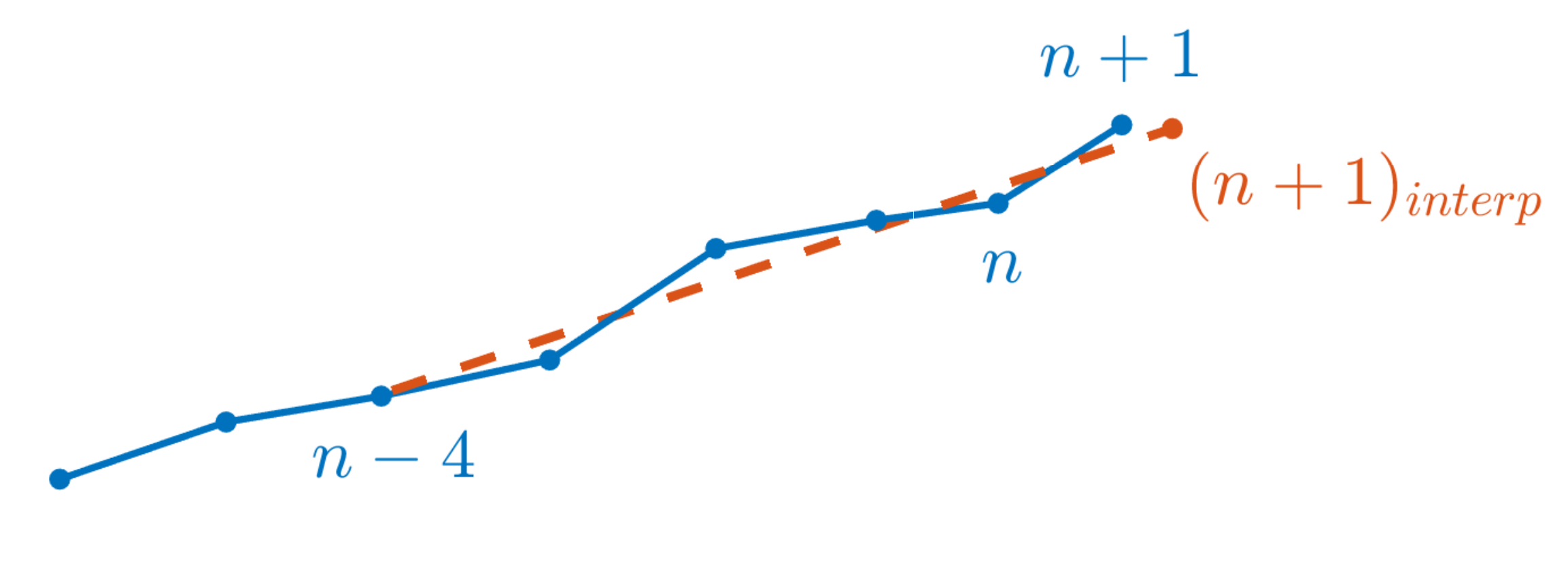}}
\caption{Predictive tracking schematic. The solid line signifies the real trajectory. The dotted line (linear fit of the positions from frame $n-4$ to $n$) indicates the position extrapolation.\label{fig:pred_track}}
\end{figure}
The stereoscopic reconstruction gives a cloud of points for every time step. The goal of the tracking is to transform this cloud into trajectories by following particles through time. To track the position of a considered particle as it moves among numerous other particles, the simplest algorithm is to consider the nearest neighbour: if one considers a particle in frame $n$, its position in frame $n+1$ is the nearest particle in frame $n+1$. But, for increased mass loading of particles, the trajectories are tangled, as observed in this study. Moreover, several points are ``ghost'' particles and should not be tracked. Thus advanced \textit{predictive tracking methods} are generally employed \citep{ouellette2006quantitative}. The trajectories are assumed to be relatively smooth and self-consistent, i.e. there are no severe variations in velocity and therefore past positions give accurate indications of future positions \citep{guezennec1994algorithms}. If one considers a particle at frame $n$, its position in frame $n+1$ can be extrapolated and finally the nearest neighbour approach is employed based on the extrapolated position. In the present study, the extrapolated position is determined by fitting the previous five positions from frame $n-4$ to $n$ with a simple linear relation (i.e. velocity), as indicated in figure~\ref{fig:pred_track}. If there are less than five positions, the available positions are used. A maximum distance of $\SI{1}{mm}$ between extrapolated position and real position is applied to continue the trajectories in order to avoid the tracking of absurd trajectories. If the same particle is the nearest neighbour for two different tracks, the nearest trajectory is chosen and the other trajectory is stopped.

\subsection{Post-processing of the trajectories}\label{subsec:postproc}
\begin{figure}
\centerline{\includegraphics[width=0.8\textwidth]{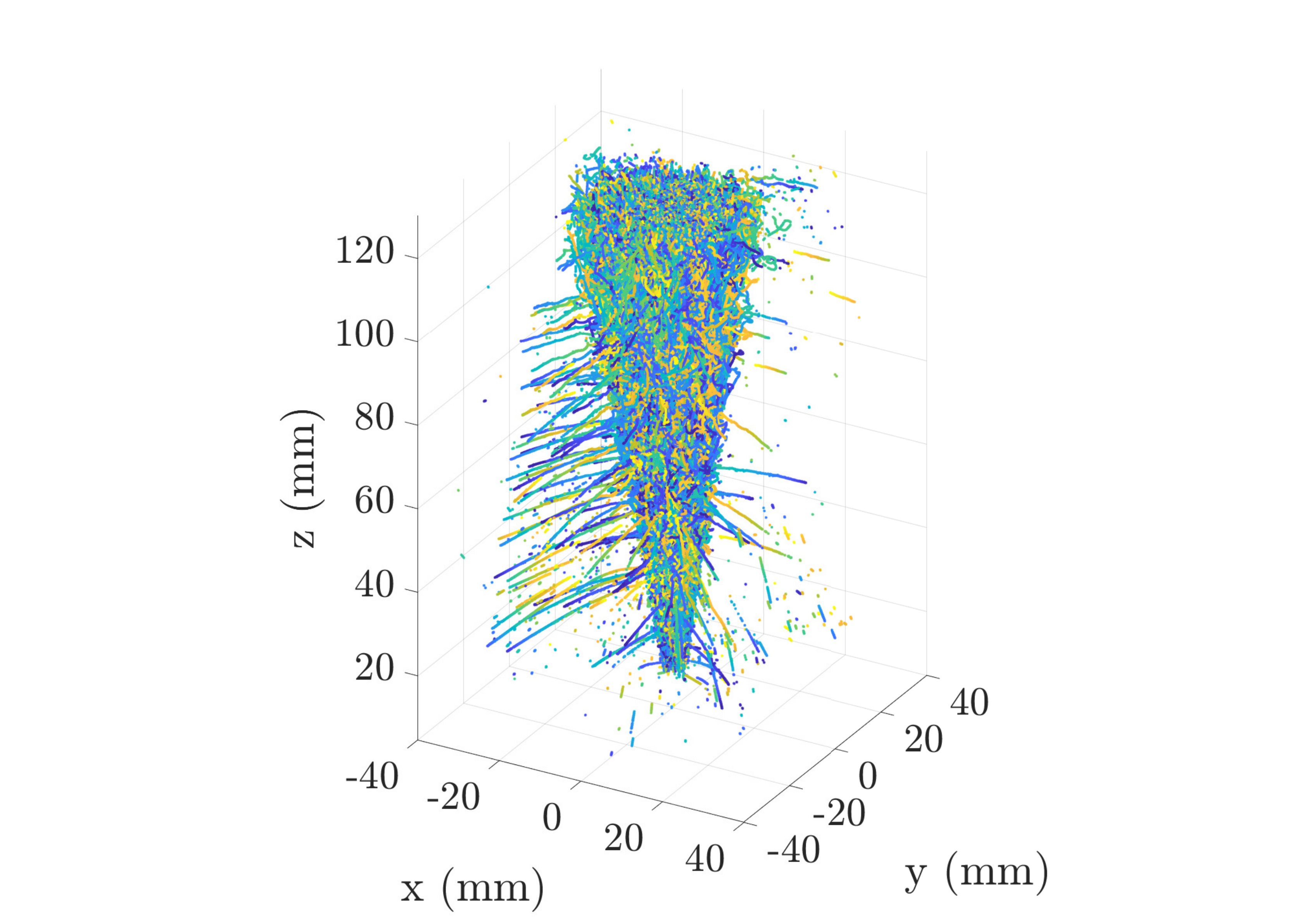}}
\caption{Near-field jet: 95 055 trajectories longer than or equal to 10 frames (one colour per trajectory, one movie considered).\label{fig:jet}}
\end{figure}
The tracking of particles results in a set of trajectories for each of the 50 experimental runs. A minimum trajectory length of 10 frames is required to remove presumably false trajectories. Some real trajectories are also removed, but their statistical value is negligible. Finally, the coordinate basis is adapted by aligning the $z$-axis with the jet axis and centring it in $x$ and $y$ directions. Positions and velocities are computed in adapted cylindrical coordinates $(z,r,\theta)$ with $z$ the axial coordinate, $r$ the radial one and $\theta$ the circumferential one. A visualisation of tracks is shown in figure~\ref{fig:jet}. It can be noted that most trajectories come from the nozzle (where they are injected) and very few come from the outside and are entrained in the jet (visible in figure~\ref{fig:jet} as radial trajectories towards the jet). The full data set for the near-field comprises $\SI{4.2e6}{}$ trajectories longer than or equal to 10 frames, which corresponds to $\SI{1.0e8}{}$ particle positions. For the far-field, it comprises $\SI{6.1e6}{}$ trajectories and $\SI{1.6e8}{}$ particle positions.

The trajectories reconstructed by the tracking algorithm always exhibit some level of noise due to errors eventually cumulated from particle detection, stereo-matching and tracking. It is important to properly handle noise, in particular when it comes to evaluating statistics associated to differentiated quantities (particle velocity and acceleration). Two techniques are implemented to do so. For all Eulerian statistical analysis requiring the estimate of local velocity, the trajectories are convolved with a first-order derivative Gaussian kernel with a length of 6 time instances and a width of 2 (\textit{ad hoc} smoothing parameters) \citep{mordant2004experimental2}. For all two-time Lagrangian statistical analysis (correlation and structure functions), an alternative noise reduction method, presented by \citet{machicoane2017estimating, machicoane2017multi}, is implemented to obtain unbiased statistics based on an estimation from discrete temporal increments of position, without requiring explicit calculation of individual trajectory derivatives. For example, to compute the noiseless Lagrangian two-point correlation of velocity, $R^L_{\hat{u}\hat{u}}$, the first-order increments are considered as follows:
\begin{equation}
R^L_{dxdx}(\tau,dt) = R^L_{\hat{u}\hat{u}}(\tau) dt^2 + \langle (db)^2 \rangle + \mathcal{O}(dt^3),
\label{eq:dt_Ruu}
\end{equation}
where $dx$ is the temporal increment of the signal $x$ over a time $dt$ with $dx = x(t+dt)-x(t) = d\hat{x}+db$. The circumflex signifies the real (noiseless) signal and the noise is denoted as $b$ (assumed to be a white noise). From the presented relationship, the noiseless correlation function of velocity $R^L_{\hat{u}\hat{u}}(\tau)$ can be extracted from the correlation of measured position increments $dx$, exploring its polynomial dependency with $dt$ at the lowest (quadratic) order and neglecting higher-order terms (i.e. $\mathcal{O}(dt^3)$), by applying a simple polynomial fit of $c_1 dt^2 + c_2$. This method, called ``$dt$-method'' in the following, allows the estimation, with increased accuracy and less sensitivity to noise, of statistics of differentiated quantities (and hence to explore small scale mechanisms), without actually requiring estimation of derivatives, but by simply considering position increments at various temporal lags. More information is provided in \citet{machicoane2017estimating, machicoane2017multi}.

\subsection{Stationarisation techniques}\label{subsec:stat}
\begin{figure}
\hspace{0.1\textwidth}(\textit{a}) \hspace{0.365\textwidth}(\textit{b})
\begin{center}
\includegraphics[height=0.264\textheight]{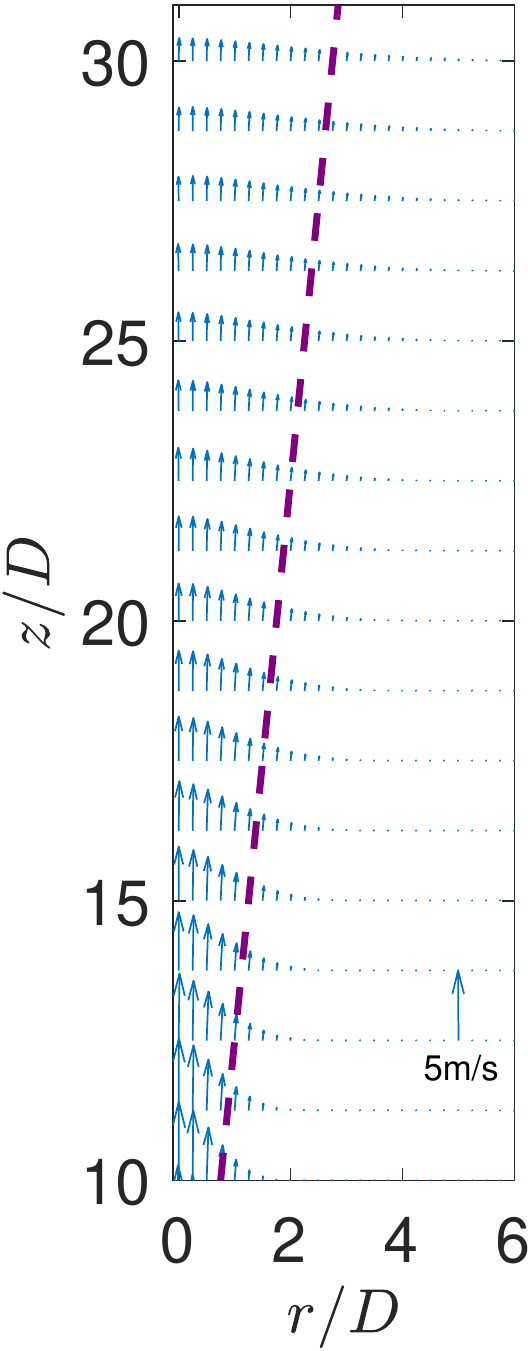}
\hspace{3cm}
\includegraphics[height=0.27\textheight]{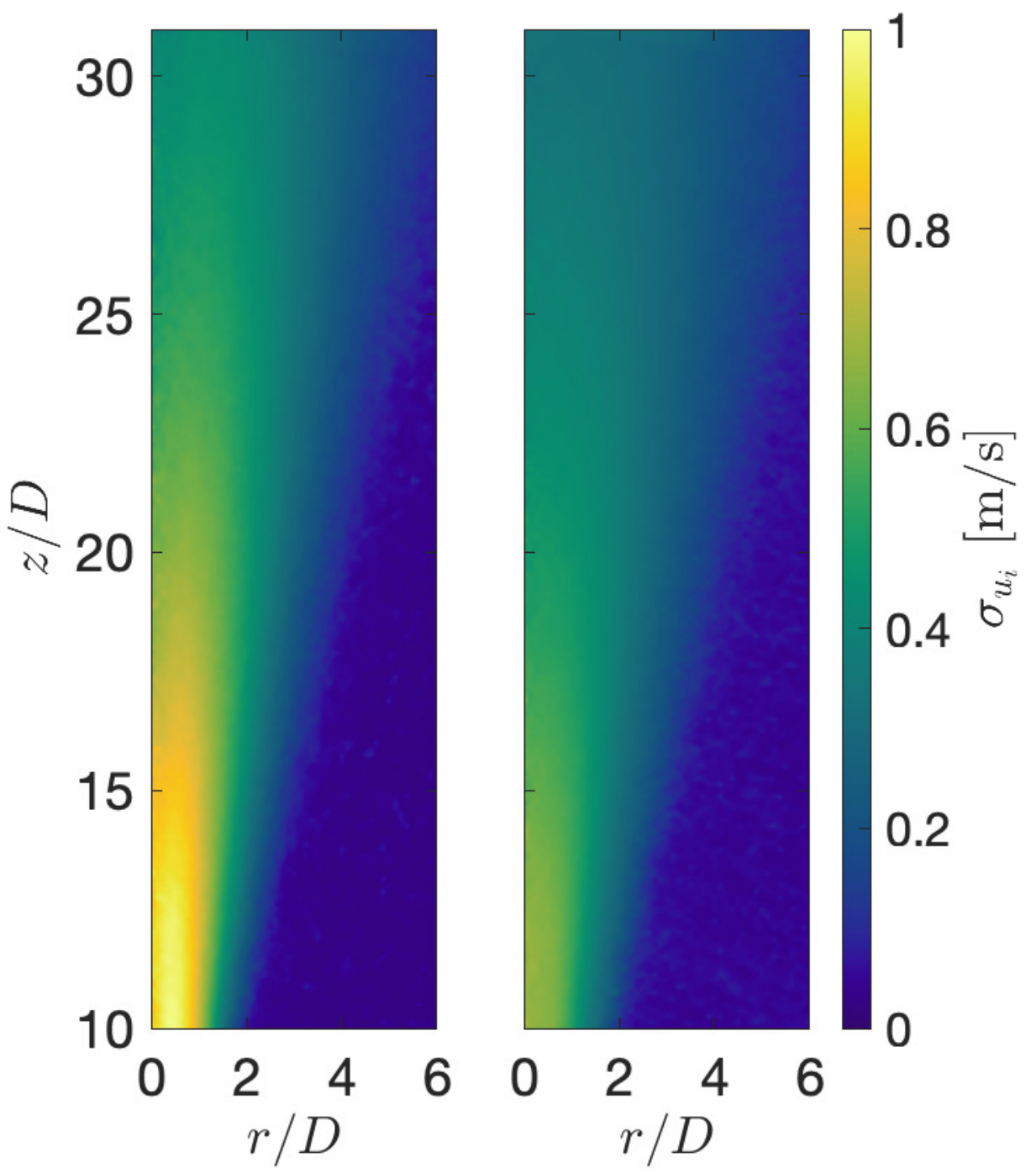}
\end{center}
\caption{(a)~Vector field of the $\boldsymbol{\overline{u}^e}$ field for the normalised locations, including the half-width of the jet $(\textcolor{purple}{--})$, $r_{1/2}$, at all downstream locations for the near-field. (b)~Contour representations of the local standard deviations $\sigma_{u_z}$ (left) and $\sigma_{u_r}$ (right) for the axial and radial velocity components for near-field locations.\label{fig:mean_fields}}
\end{figure}
To address the Lagrangian instationarity (related to the Eulerian inhomogeneity) of the flow, methods are taken according to the proposed self-similarity of a turbulent jet by \citet{batchelor1957diffusion}, i.e. based on the transformation of the Lagrangian velocity and time scales of a particle at a given time $\tau$ after it has been released from a point source. Equation~\eqref{eq:ucomp} provides a relationship to achieve proper stationarisation. For this study, the fluctuating stationarised velocity is obtained by subtracting the local Eulerian velocity (and assuming cylindrical symmetry of the jet, hence neglecting the $\theta$ dependence on spatially averaged quantities), $\overline{u}^e_i(z,r)$, and scaled by the local standard deviation, $\sigma_{u_i}(z,r)$. Explicitly,
\begin{equation}
\tilde{u}_i(\tau) = \frac{u_i(\tau) - \overline{u}^e_i(\boldsymbol{x}(\tau))}{\sigma_{u_i}(\boldsymbol{x}(\tau))} = \frac{u_i(\tau) - \overline{u}^e_i(z,r)}{\sigma_{u_i}(z,r)}.
\label{eq:stat}
\end{equation}
The local standard deviation is an optimal choice for compensation as it generalises the methods presented in \citet{batchelor1957diffusion}, where a specific decay rate (Batchelor assumed a power-law) is required for stationarisation. This velocity $\tilde{u}$ takes the mean drift and decay into account although the term becomes dimensionless as a result. For this reason, for all statistical calculations of dimensional quantities (such as the turbulent dissipation rate) inferred from this analysis, velocity is redimensionalised through multiplication with the average local standard deviation within the considered measurement region or location. For transparency, the Eulerian mean and standard deviation velocity fields used for the stationarisation are presented in figure~\ref{fig:mean_fields} (figure~\ref{fig:mean_fields}(a) the mean velocity as a vector field and figure~\ref{fig:mean_fields}(b) the standard deviation of the axial and radial velocity components). The half-width of the jet, $r_{1/2}(z)$, where $\overline{u}^e_z(z,r=r_{1/2}(z)) = \thalf \overline{u}^e_z(z,r=0)$, is included in the Eulerian mean velocity field as the dashed line to provide clarity to the sampling methods based on this quantity, as discussed in sections~\ref{sec:eul} and~\ref{sec:lag}. Note that Lagrangian velocity components are used for the Eulerian statistical characterisation therefore the stationarisation technique described is required for all analyses presented in the study. For clarity, herein the tilde is omitted and the compensated Lagrangian velocity is denoted as $u(\tau)$ for the remainder of the article.

\section{Eulerian velocity statistical analysis}\label{sec:eul}
This section aims to extract flow parameters such as length scales and energy dissipation rate from different Eulerian statistics: second-order structure functions and two-point correlation functions. The jet flow is inhomogeneous, therefore these quantities depend on $z$ and $r$. Focus is placed on centerline statistics for the Eulerian characterisation of the jet, limited to radial distances up to $r_{1/2}$ and consideration of only the $z$-axis evolution is used to characterise the main property of the base turbulence.

\subsection{Eulerian second-order structure functions}
Structure functions are commonly used to describe multi-scale properties of turbulence through a statistical representation of a flow quantity with a given spatial or temporal separation. In the Eulerian perspective, longitudinal velocity structure functions of order $n$ are defined as
\begin{equation}
S^E_{n-\parallel}(\boldsymbol{x},\boldsymbol{\Delta x}) = \langle [\delta u_\parallel(\boldsymbol{x},\boldsymbol{\Delta x})]^n \rangle = \langle [u_\parallel(\boldsymbol{x}+\boldsymbol{\Delta x}) - u_\parallel(\boldsymbol{x})]^n \rangle,
\end{equation}
where $\delta u_\parallel$ is computed over two points, one at $\boldsymbol{x}$, the other at $\boldsymbol{x}+\boldsymbol{\Delta x}$, with $u_\parallel$ defined as the single longitudinal component of Eulerian velocity along $\boldsymbol{\Delta x}$. The $\langle \cdot \rangle$ denotes ensemble averaging.

In homogeneous isotropic stationary turbulence (HIST), Kolmogorov phenomenology K41 \citep{kolmogorov1941local} predicts for the second-order structure function in the \textit{inertial range}, scales between the Kolmogorov scale, $\eta$, and the integral length scale, $L$, that: 
\begin{equation}
S^E_{2-\parallel}(\Delta) = \langle [\delta u_\parallel(\boldsymbol{x},\Delta)]^2 \rangle = C_2 \frac{(\varepsilon \Delta )^{2/3}}{\sigma_{u_\parallel}^2},
\label{eq:S2_K41}
\end{equation}
with $\varepsilon$ the average energy dissipation rate per unit mass and $C_2 \simeq 2.0$ \citep{pope2000turbulent}. The $\sigma_{u_\parallel}^2$ denominator (the variance of longitudinal velocity component) has been added here in the right hand term to account for the fact that the stationarised velocity according to transformations~\eqref{eq:stat} is considered. Alternatively, the transverse structure function $S^E_{2-\perp}(\Delta)$ can be considered where increments are taken for the velocity components perpendicular to the separation vector. In HIST, within the inertial range, $S^E_{2-\perp}(\Delta)$ follows the same K41 scaling but with a constant $C_{2\perp} = \frac{4}{3} C_2$. Previous studies have found that these relations \textit{a priori} established for HIST, apply reasonably well to the inertial scales of turbulent jets, in spite of the large scale inhomogeneity and anisotropy (see for instance \citet{romano2001longitudinal}). In the sequel relation~\eqref{eq:S2_K41} is used together with the relation $C_{2\perp}=\frac{4}{3}C_2$ to analyse longitudinal and transverse structure functions in the jet.

Within the jet (cylindrical coordinates), the longitudinal second-order structure function is usually estimated, near the centerline, based on the axial component of the velocity:
\begin{equation}
S^E_{2-z,\parallel}(z,\delta z) = \langle [u_z(z+\delta z,r) - u_z(z,r)]^2 \rangle,
\end{equation}
with $u_z$ the fluctuating axial velocity (recall that the stationarisation described in section~\ref{subsec:stat} is applied) and $\delta z$ the axial distance between the two considered points (the explicit $z$ dependency is kept here to emphasise the streamwise inhomogeneity of the jet centerline statistics). This is, for instance, the quantity typically measured when using hot-wire anemometry (sensitive to the streamwise velocity component) combined with the Taylor frozen field hypothesis.

To explore the streamwise evolution of Eulerian properties of the jet, a set of data (particle velocities) is considered for a given $z$ position, which falls within a short cylinder (\textit{disk}), $\mathcal{D}_z$, of limited height ($\SI{0.5}{mm}$ in the $z$-direction) and a radius of $r_{1/2}(z)$ for statistical analysis. The disk radius is chosen to include sufficient particles for statistical convergence but, in being limited to the half-width, the volume does not encompass particles from the turbulent/non-turbulent interface. This gives a canonical description of turbulent properties representative of the centerline of the jet. Consideration of statistics in a thin disk allows the more detailed exploration of $z$ dependence of statistical quantities, however this sampling technique forbids exploration of $\delta z$ values over a range relevant to estimate $S^E_{2-z,\parallel}(z,\delta z)$ at inertial scales. To overcome this issue, two strategies are considered: (i)~Still based on the axial $z$-component of the velocity, $S^E_{2-z,\perp}(z,\delta r)$, the transverse structure function of $u_z$ (with the separation vector $\boldsymbol{\delta r}$ taken within the plane of the disk) is estimated in lieu of $S^E_{2-z,\parallel}(z,\delta z)$; (ii)~For radial velocities, the longitudinal structure function is considered through use of the velocity components perpendicular to the $z$-axis (i.e. within the sampling disk $\mathcal{D}_z$), projected onto the increment vector $\boldsymbol{\delta r}$ within the disk $\mathcal{D}_z$. This is denoted as $S^E_{2-r\theta,\parallel}(z,\delta r)$ (where the subscript $r\theta$ recalls that only velocity components perpendicular to $z$ are considered). For any redimensionalization of a Eulerian quantity, the averaged standard deviation within a respective disk, $\langle \sigma_{u_i} \rangle_{\mathcal{D}_z}$, is employed. For brevity this is herein denoted as $\sigma_{u_i}$ for all Eulerian calculations.

The discussions of this subsection (and in the two following) illustrate the extraction of the main Eulerian turbulent properties (and of their streamwise evolution) based on $S^E_{2-z,\perp}(z,\delta r)$. The same analysis was also repeated based on $S^E_{2-r\theta,\parallel}(z,\delta r)$, the details of which are not provided for brevity. Analysis follows the same recipe as is described for $S^E_{2-z,\perp}(\delta r)$, and the main extracted turbulent parameters from these two estimates will be discussed and compared in section~\ref{subsec:eulParam}.

The transverse structure function based on $u_z$ at a given $z$ position is estimated as 
\begin{equation}
S^E_{2-z,\perp}(z,\delta r) = \langle [u_z(\boldsymbol{r}+\boldsymbol{\delta r}) - u_z(\boldsymbol{r})]^2 \rangle_{\mathcal{D}_z},
\label{eq:S2_z}
\end{equation}
where the average is taken over pair of particles within the disk $\mathcal{D}_z$ separated by a vector $\boldsymbol{\delta r}$. Note that, given the reduced height of the disk (not exceeding two particle diameters), $\boldsymbol{\delta r}$ is within an acceptable approximation perpendicular to the $z$ axis, ensuring that equation~\eqref{eq:S2_z} indeed corresponds to a transverse structure function (except maybe for the smallest separations, comparable to the disk height).

\begin{figure}
\hspace{0.02\textwidth}(\textit{a}) \hspace{0.445\textwidth}(\textit{b})\\
\centerline{\includegraphics[width=0.95\textwidth]{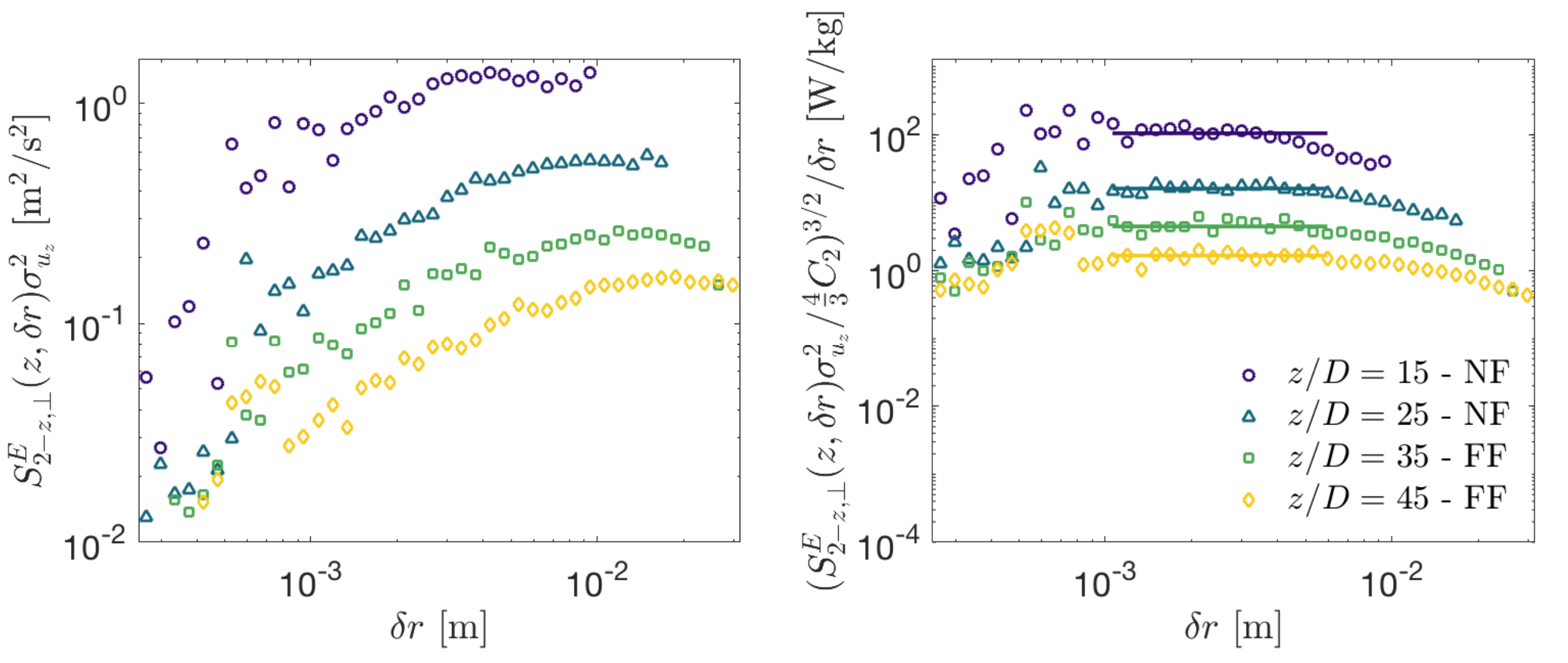}}
\caption{Eulerian second-order structure functions of the axial velocity on the axis, (a)~uncompensated $S^E_{2-z,\perp}(z,\delta r)\sigma_{u_z}^2$ and (b)~compensated $(S^E_{2-z,\perp}(z,\delta r)\sigma_{u_z}^2/\frac{4}{3}C_2)^{3/2}/\delta r$ (the solid lines are the plateaus to extract $\varepsilon_z$), for the four denoted downstream locations. \label{fig:s2_eul}}
\end{figure}
$S^E_{2-z,\perp}(z,\delta r)$ is computed for different $z$ positions (in the near and far-fields of the jet) and shown in figure~\ref{fig:s2_eul}(a). As explained in section~\ref{subsec:stat}, while the stationarised (hence dimensionless) velocity is used for all estimates, $S^E_{2-z,\perp}$ is made dimensional by multiplying it by the square of $\sigma_{u_z}$, the standard deviation of $u_z$ within $\mathcal{D}_z$ (see table~\ref{tab:eul_parameters}). This redimensionalisation is required in order to extract the dimensional value of $\varepsilon$, and the associated derived parameters (in particular the dissipation scales and Taylor micro-scale). To this end, figure~\ref{fig:s2_eul}(b) includes the compensated structure function $(S^E_{2-z,\perp}(z,\delta r)\sigma_{u_z}^2/\frac{4}{3}C_2)^{3/2}/\delta r$ (measurements by \citet{romano2001longitudinal} suggest that at in spite of the large scale anisotropy, the isotropic relation $C_{2\perp} = \frac{4}{3}C_2$ applies reasonably well for the inertial scale dynamics of the jet). Well defined plateaus, corresponding to inertial range dynamics, are observed from which the dissipation rate $\varepsilon_z$ can be extracted according to equation~\eqref{eq:S2_K41}. The subscript $z$ in $\varepsilon_z$ simply refers to the fact that this estimate is based on the axial component of the velocity. It will be compared later with $\varepsilon_{r\theta}$, the estimate from $S^E_{2-r\theta,\parallel}$. It can be seen that, as the location downstream increases, the plateau of the second-order structure function (and hence $\varepsilon_z$) decreases, due to the streamwise decay of turbulence along the jet.

It is noted that small scales (typically for $\delta r < \SI{e-3}{m}$) are not statistically well converged. This is due to the lack of statistics for pairs of particles with very small separation due to the moderate seeding of particles used for the Lagrangian tracking.

\subsection{Eulerian two-point correlation functions}
\begin{figure}
\centerline{\includegraphics[width=0.5\textwidth]{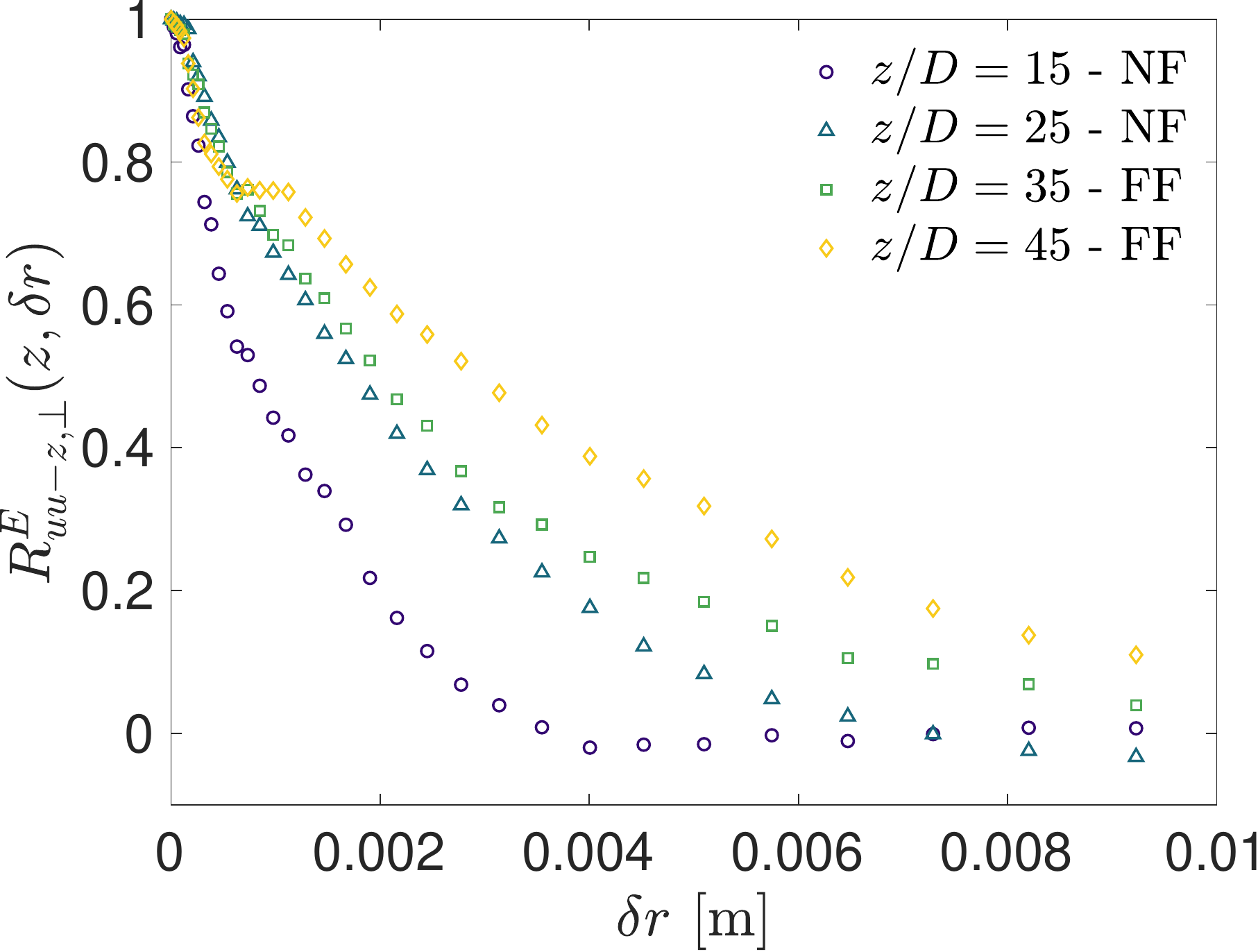}}
\caption{Normalised two-point spatial correlation of the Eulerian axial velocity on the axis, $R^E_{uu-z,\perp}(z,\delta r) = 1 - S^E_{2-z,\perp}(z,\delta r)/2 $. \label{fig:ruu_eul}}
\end{figure}
The second-order Eulerian statistics shown in the previous section from the structure functions can be equivalently investigated in terms of the two-point correlation function. The correlation of axial velocity can indeed be obtained via the non-dimensional second-order structure function, $R^E_{uu-z,\perp}(z,\delta r) = 1 - S^E_{2-z,\perp}(z,\delta r)/2$, to depict the evolution of the velocity interactions through space. The results from the near-field and far-field are presented in figure~\ref{fig:ruu_eul}. The curves are ordered depending on their downstream location $z$. The location nearest the jet exit, $z/D = 15$, exhibits a rapid decorrelation. As the flow advances downstream, the turbulent length scales grow, resulting in a dynamics which remains correlated over longer distances, as seen by the $z/D = 45$ profile. This trend can be investigated quantitatively using the Eulerian correlation length (or Eulerian integral scale) $L_{E_{z,\perp}}(z) = \int_0^\infty R^E_{uu-z,\perp}(z,\delta r) \:\mathrm{d}\delta r$. Recall that transverse and longitudinal correlation lengths are kinematically related in HIST by $L_{E_{z,\parallel}} = 2 L_{E_{z,\perp}}$ \citep{pope2000turbulent}. Since most studies in the literature refer to the longitudinal length, the present study will then consider $L_{E_z}(z) = 2 \int_0^\infty R^E_{uu-z,\perp}(z,\delta r) \:\mathrm{d}\delta r$, avoiding the $\perp$ or $\parallel$ subscripts. However, it is noted that measurements by \citet{burattini2005similarity} suggest that the ratio may actually be slightly lower than 2, and closer to 1.8 in free shearing jets due to large scale anisotropy.

\subsection{Evolution of Eulerian parameters}\label{subsec:eulParam}
\begin{figure}
\centerline{\includegraphics[width=0.8\textwidth]{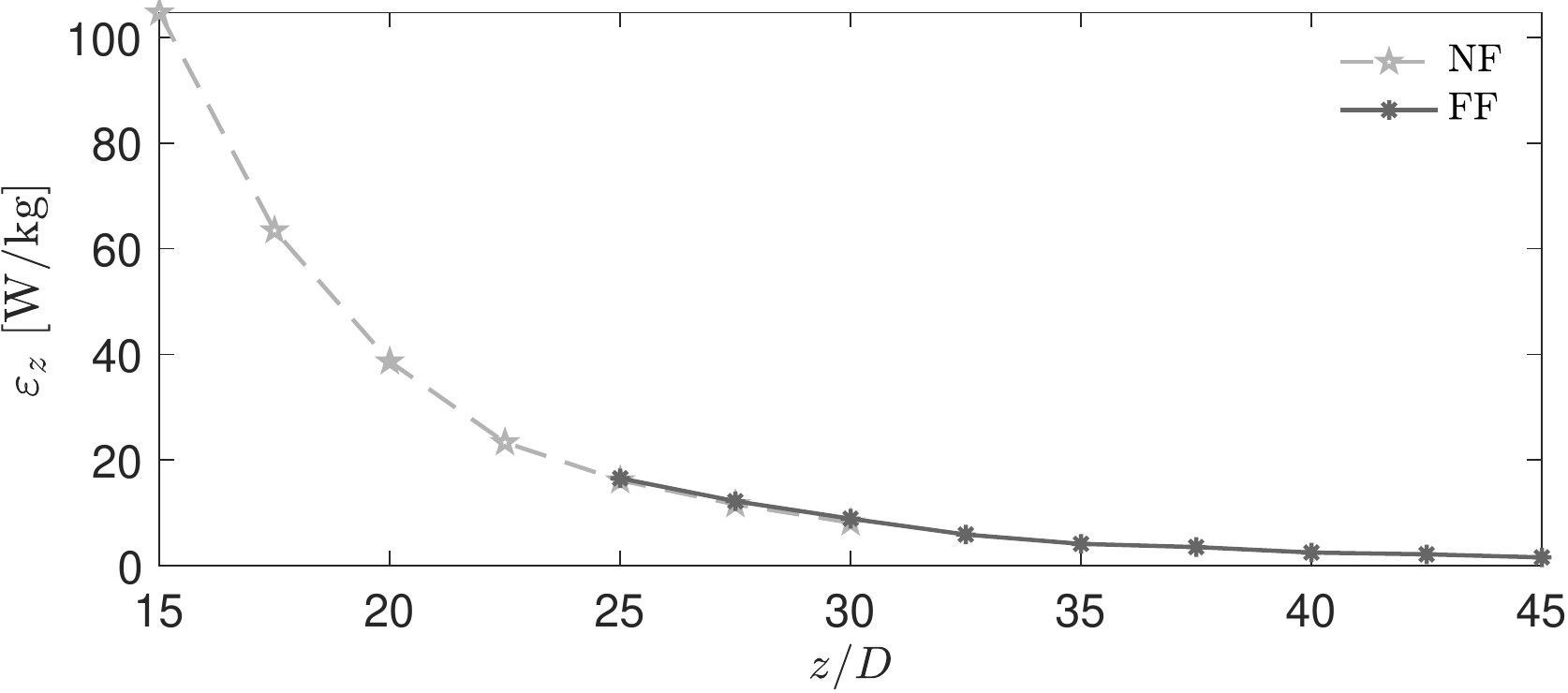}}
\caption{Evolution of $\varepsilon_z$ along the jet axis.\label{fig:epsilon}}
\end{figure}
\begin{figure}
\hspace{0.15\textwidth}(\textit{a}) \hspace{0.22\textwidth}(\textit{b}) \hspace{0.22\textwidth}(\textit{c})\\
\centerline{\includegraphics[width=0.75\textwidth]{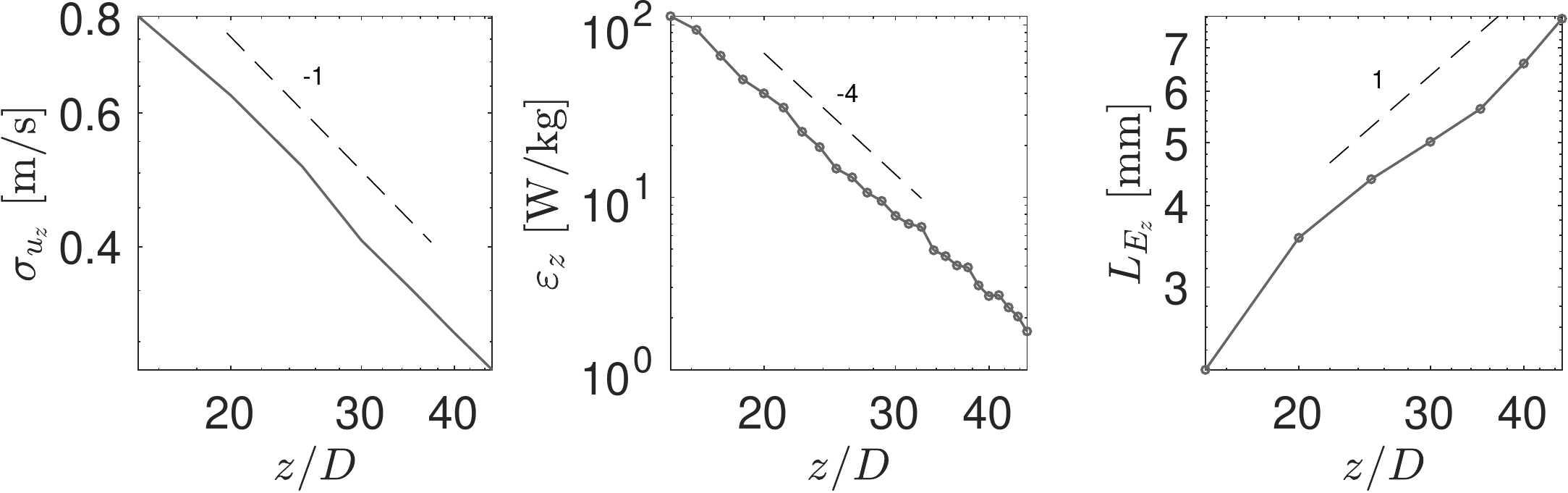}}
\caption{(a)~The standard deviation averaged within the disk $\mathcal{D}_z$, (b)~the dissipation rate and (c)~the integral length scale for the axial component of velocity for all downstream locations. Power-law relation is given as dashed line. 
\label{fig:4graphs}}
\end{figure}

The evolution of $\varepsilon_z$, estimated from the plateaus of the compensated second-order structure functions (figure~\ref{fig:s2_eul}(b)), is represented in figure~\ref{fig:epsilon}. There exists a tendency of $1/z^4$ (more clearly visible in figure~\ref{fig:4graphs}(c)), as expected for canonical self-similar jets. The observed consistency in the values and shape of the profiles between the near and far-field experimental locations, validates the presented $\varepsilon_z$ values from the independent measurements carried over the overlapping region.

\begin{table}
\centerline{\scalebox{1}{
\begin{tabular}{ccccccccc|cccc}
$z/D$ & $\sigma_{u_z}$ & $\varepsilon_z$ & $\eta_z$ & $\tau_{\eta_z}$ & $\lambda_z$ & $\Rey_\lambda$ & $L_{E_z}$ & $T_{E_z}$ & $\sigma_{u_r}$ & $\varepsilon_{r\theta}$ & $L_{E_{r\theta}}$ & $T_{E_{r\theta}}$\\[1pt]
& $[\SI{}{m/s}]$ & $[\SI{}{W/kg}]$ & $[\SI{}{\micro\meter}]$ & $[\SI{}{ms}]$ & $[\SI{}{\micro\meter}]$ & & $[\SI{}{mm}]$ & $[\SI{}{ms}]$ & $[\SI{}{m/s}]$ &$[\SI{}{W/kg}]$ & $[\SI{}{mm}]$ & $[\SI{}{ms}]$\\[3pt]
15 & 0.80 & 104.7 & 9.9 & 0.098 & 304 & 245 & 2.2 & 2.8 & 0.57 & 63.9 & 1.7 & 2.9\\
25 & 0.51 & 16.1 & 15.8 & 0.249 & 491 & 250 & 4.4 & 8.6 & 0.38 & 14.6 & 2.0 & 5.3\\
35 & 0.35 & 4.5 & 21.7 & 0.472 & 643 & 226 & 5.6 & 16.0 & 0.28 & 5.7 & 3.6 & 13.0\\
45 & 0.28 & 1.7 & 27.8 & 0.774 & 825 & 227 & 7.8 & 28.2 & 0.22 & 2.4 & 5.1 & 23.2\\
\end{tabular}}}
\caption{Eulerian parameters of the jet on the axis for different $z/D$ positions. \label{tab:eul_parameters}}
\end{table}
From the dissipation rate $\varepsilon_z$, other relevant parameters of the flow field can be extracted, namely the Kolmogorov time scale, $\tau_{\eta_z} = (\nu/\varepsilon_z)^{1/2}$, and length scale, $\eta_z = (\nu^3/\varepsilon_z)^{1/4}$, as well as the Taylor microscale, $\lambda_z = (15\nu\sigma_{u_z}^2/\varepsilon_z)^{1/2}$, and the Taylor-based Reynolds number $\Rey_\lambda = \sigma_{u_z}\lambda_z/\nu$, both of which assume HIST. Further, large length and time scales are obtain from the two-point correlation profiles in figure~\ref{fig:ruu_eul}. For a more accurate estimate of the correlation length, $L_{E_z}(z) = 2 \int_0^\infty R^E_{uu-z,\perp}(z,\delta r) \:\mathrm{d}\delta r$, the integral of the correlation functions is based on a fit of the curves shown in figure~\ref{fig:ruu_eul} using a Batchelor type parametrisation \citep{lohse1995bottleneck}. Recall that the factor 2 is the HIST correction that relates the transverse correlation (given by the integral of $R^E_{uu-z,\perp}$) to the longitudinal one. The calculated $L_{E_z}$ shall therefore be interpreted as the longitudinal integral scale associated to the $z$ component of velocity. The integral time scale is then computed as $T_{E_z}=L_{E_z}/\sigma_{u_z}$. All relevant quantities of the jet have been accumulated for the considered streamwise locations in table~\ref{tab:eul_parameters}. The streamwise evolution for the velocity standard deviation, dissipation rate and integral scale are also shown in figure~\ref{fig:4graphs}, where the well known self-similar power-law profiles can be seen.

This brief characterisation of basic Eulerian properties is concluded by reporting the similarly employed analysis performed on $S^E_{2-r\theta,\parallel}(z,\delta r)$ (where rather than $u_z$, components of velocity perpendicular to the z-direction are considered). This leads to estimates of the dissipation rate $\varepsilon_{r\theta}$ (and derived quantities) and of the integral length scale and time scale, $L_{E_{r\theta}}$ and $T_{E_{r\theta}}$. Results are included in table~\ref{tab:eul_parameters}. The dissipation rate for the radial velocity is lower than the axial component near the exit of the jet, but declines more slowly as the jet develops, resulting in similar values for $\varepsilon_{r\theta}$ and $\varepsilon_z$ at $z/D > 25$. As a result, in this region dissipation scales are found almost identical with both estimates. This supports the idea that small and inertial scales are nearly isotropic. The large scales show however a certain degree of anisotropy, in particular regarding the integral length scale and in a lesser degree the integral time scale which are found larger for the $z$-component than for the $r\theta$-components. 

\section{Lagrangian velocity statistical analysis}\label{sec:lag}
In this section the Lagrangian statistics of the jet dynamics are investigated with a particular focus on second-order statistics (namely velocity second-order structure function and two-point correlation function), which are key ingredients to model turbulent diffusion, as discussed in the introduction. In particular, the relevance of Batchelor's Lagrangian self-similar stationarisation idea is further assessed.

\subsection{Lagrangian second-order structure functions}\label{subsec:lagS2}
\begin{figure}
\centerline{\includegraphics[width=0.5\textwidth]{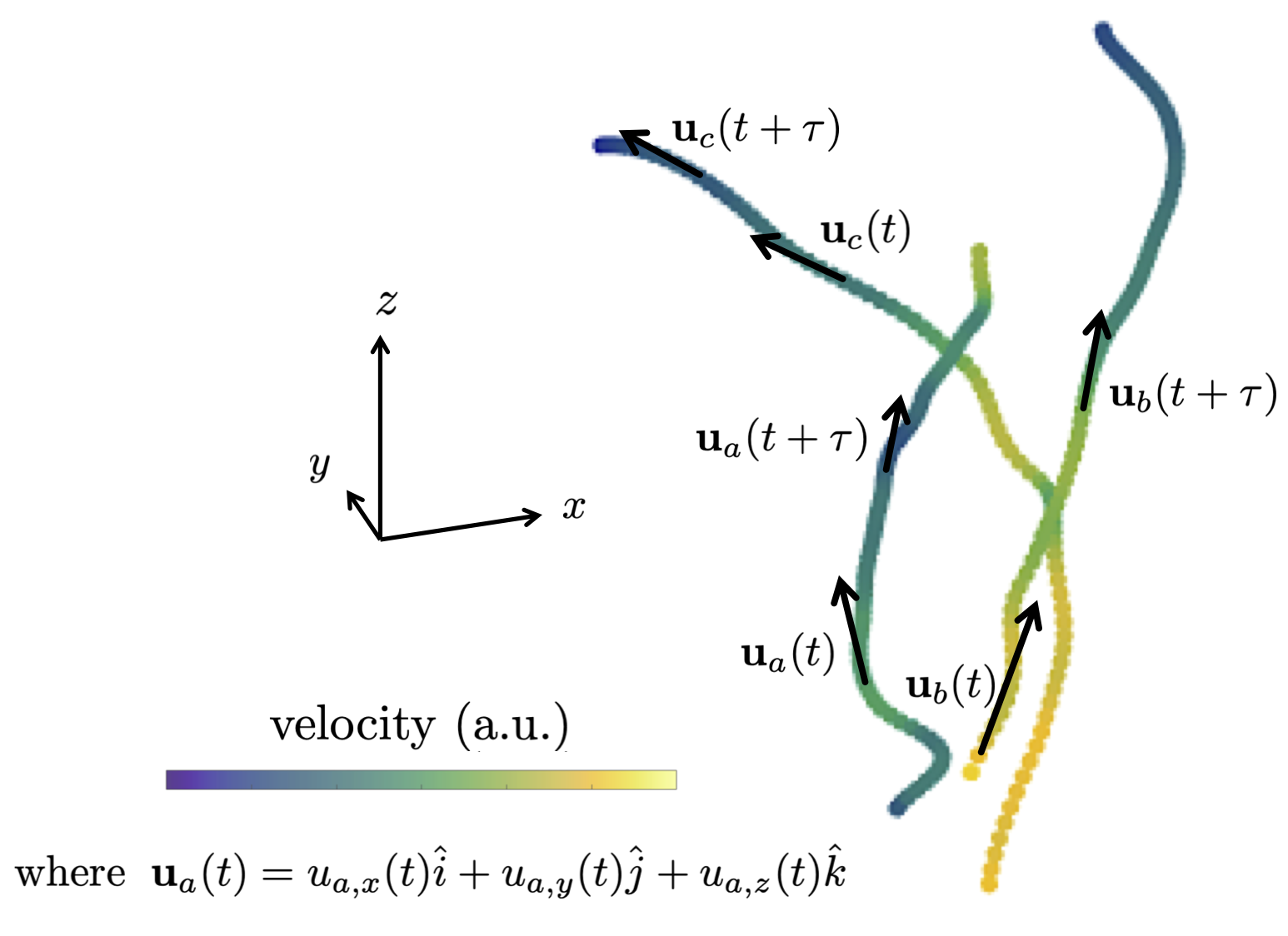}}
\caption{Schematic of the Lagrangian velocity increment in a Cartesian coordinate system for a given time lag $\tau$.\label{fig:vel_inc}}
\end{figure}
The application of the known Kolmogorov K41 phenomenology for HIST, generally applied to Eulerian inertial scaling, can be extended to the Lagrangian framework \citep{toschi2009lagrangian}, where dynamics is investigated as a function of temporal increments along particle trajectories (see figure~\ref{fig:vel_inc}). Namely, for the second order Lagrangian structure function, this reads (for the stationarised velocity defined by relation~\eqref{eq:stat}):
\begin{equation}
S^L_{2,i}(\tau) = \langle [u_i(t+\tau) - u_i(t)]^2 \rangle = C_{0_i} \frac{\varepsilon_i \tau}{\sigma_{u_i}^2},
\label{eq:s2_lag}
\end{equation}
within the the inertial range ($\tau_\eta \ll \tau \ll T_L$), where $i$ is a velocity component ($i=x$, $y$ or $z$, by symmetry, statistics along $x$ and $y$ are identical and equivalent to statistics of the radial $r$-component of velocity), and $T_L$ is the Lagrangian integral time scale, which is expected to be related to the Eulerian integral time (this point will be further deepen later). Note that while for the Eulerian structure functions, spatial velocity increments were computed between pairs of particles and then averaged, now, for Lagrangian analysis, temporal velocity increments are computed on each individual trajectories before being averaged. In this study, the nature of this scaling is revisited as well as the value of the constant $C_0$ when the stationarised velocity presented in section~\ref{subsec:stat} is considered. In order to address the role of jet anisotropy (in particular regarding the value of $C_{0_z}$ and $C_{0_r}$), the statistics for the axial and radial components of velocity are considered separately. 
Recall also that in lieu of Gaussian filtering previously applied to the Eulerian structure functions, the $dt$-method presented in section~\ref{subsec:postproc} is implemented, which has been shown to better handle noise issues for Lagrangian velocity statistics estimates \citep{machicoane2017estimating}.

\begin{figure}
\hspace{0.02\textwidth}(\textit{a}) \hspace{0.445\textwidth}(\textit{b})\\
\centerline{\includegraphics[width=0.95\textwidth]{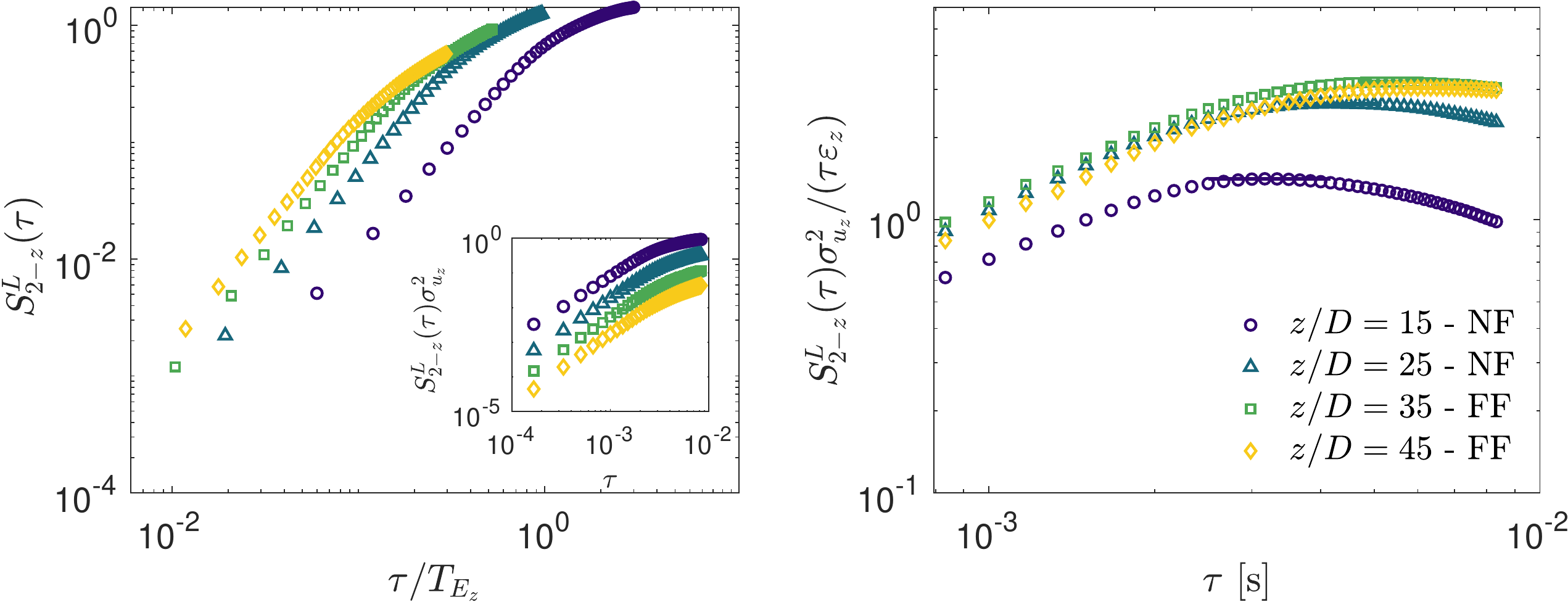}}
\caption{Lagrangian second-order structure functions of the axial velocity on the axis, estimated at four downstream locations ($z/D =$ 15, 25, 35, and 45). (a)~Non-dimensional $S^L_{2,z}(\tau)$ as a function of the non-dimensional time $\tau/T_{E_z}$ (dimensional $S^L_{2,z}(\tau)\sigma_{u_z}^2$ as a function of time $\tau$ in inset) and (b)~compensated $S^L_{2,z}(\tau)\sigma_{u_z}^2/(\tau\varepsilon_z)$, for the denoted downstream locations. The universal scaling constant, $C_{0_z}$, can be extracted from the plateau of the compensated structure functions.\label{fig:s2_lag}}
\end{figure}
Figure~\ref{fig:s2_lag}(a) presents the corresponding curves for $S^L_{2,z}(\tau)$ at the four different downstream locations. For each location $z$, the ensemble selected for the Lagrangian statistics consists of all trajectories passing through a small sphere, $\mathcal S_z$ centred at downstream position $z$ along the jet centerline, with a radius of $r_{1/2}(z)/3$. This volume allows sufficient particles for convergence of statistics yet does not overlap in the axial direction as the half-width increases. Similar to methods presented in the Eulerian framework, the averaged standard deviation from within each respective sphere, $\langle \sigma_{u_i} \rangle_{\mathcal{S}_z}$, is used for redimensionalization of Lagrangian quantities when necessary (for calculation of $C_0$) and denoted simply as $\sigma_{u_i}$.  All curves exhibit a transition from a dissipative behaviour at small time lags (where $S^L_{2,z} \propto \tau^2$) to the inertial range (where $S^L_{2,z} \propto \tau$). The main figure shows the structure function in stationarised variables, while the inset provides the same data but non-stationarised. Several interesting points emerge:

\begin{itemize}
\item \textbf{Effect of stationarisation at inertial scales.} The non-stationarised statistics (inset of figure~\ref{fig:s2_lag}(a) are widely spread while the stationarised statistics (main figure) collapse reasonably well, in particular in the far-field ($z/D > 25$). Similarity between the curves is improved for the inertial range dynamics (which presents similar trends even at distances $z/D \gtrsim 20$), but less adequate for the small scale dissipative dynamics, for which the collapse becomes reasonable only at far downstream locations ($z/D > 35$). This suggests that the stationarisation procedure is efficient to retrieve self-similar inertial range Lagrangian statistics in the far-field (in Batchelor's sense, meaning that Lagrangian statistics become independent of the downstream position as particles travel along the jet), while discrepancies remain in the small scales until the very far-field.

\item \textbf{Small scale dynamics discrepancies.} In the Lagrangian framework, the small scale dynamics of structure functions is associated to particle acceleration statistics. Figure~\ref{fig:s2_lag}(a) therefore suggests that stationarised acceleration statistics eventually fall in line, but only in the very far-field (curves at $z/D = 40$ and 45 almost perfectly collapse). As will be observed in section~\ref{sec:acc}, acceleration statistics are strongly affected by the finite size of the particles, which in our study remains much larger than the dissipation scale of the flow ($d_p/\eta = 25$ at $z/D = 15$ and 9 at $z/D = 45$). Although further investigation focusing specifically on the small scale dynamics would be required (which is not the scope of the present article, mostly motivated by applications to diffusion which is primarily driven by inertial and large scale behaviour), it is probable that the observed discrepancy at small scales reflects these finite size effects. This is supported by the fact that as considered positions are farther downstream (where $d_p/\eta$ gets smaller and hence finite particle size effects disappear), the stationarised acceleration dynamics seems to better converge to a single curve. Accelerations statistics and finite size effects will be further discussed in section~\ref{sec:acc}.

\item \textbf{Large scales dynamics.} By construction, the second order structure function of the stationarised velocity should reach, in the large scales, an asymptotic constant value of 2 as the Lagrangian dynamics becomes fully decorrelated. This asymptotic regime is not reached in our data, where $S^L_2$ reaches at best values of order 1, without exhibiting an asymptotic decorrelated plateau. This is due to the lack of statistics for long trajectories. One of the well-known difficulties of Lagrangian diagnosis is indeed the capacity to obtain sufficiently long trajectories allowing to explore the large scale dynamics. In the present study, most trajectories are efficiently tracked over a few tens of frames at most (very few are over hundreds of frames). At the operating repetition rate of 6000 frames per second, this corresponds to trajectories at most $\SI{10}{ms}$ long, what represents (according to table~\ref{tab:eul_parameters}) a few Eulerian integral times scales in the near-field, and only a fraction of this integral scale in the far-field, where only a part of the inertial range dynamics is accessible. In subsection~\ref{subsec:lagautocorr}, it is demonstrated that large scale behaviour (and the effect of stationarisation on it), can still be addressed by estimating the Lagrangian correlation time scales.

\item \textbf{Estimate of $C_0$ constant.} Figure~\ref{fig:s2_lag}(b) shows the compensated structure functions, $S^L_{2,z}(\tau)\sigma_{u_z}^2/(\tau \varepsilon_z)$, built with $\varepsilon_z$ values found in the Eulerian analysis (for consistency regarding possible anisotropy effects, the estimate of dissipation rate based on Eulerian statistics of corresponding components is used). Based on relation~\eqref{eq:s2_lag}, within the inertial range the value of $C_{0_z}$ can be extracted from the plateau of the curves. The value of the plateau is observed to saturate, as considered positions reach farther towards the far-field, at a value of $C_{0_z} \simeq 3.2$. The downstream evolution of $C_{0_z}$ will be further discussed in the coming sections.

\end{itemize}

All observations also apply to estimates of $S^L_{2,r}$, based on the radial component of velocity. Quantitative comparison of the downstream evolution of $C_{0_z}$ and $C_{0_r}$ will be detailed in section \ref{subsec:lagParam}.

\subsection{Lagrangian two-point correlation functions}\label{subsec:lagautocorr}
\begin{figure}
\hspace{0.03\textwidth}(\textit{a}) \hspace{0.38\textwidth}(\textit{b})
\begin{center}
\includegraphics[height=0.18\textheight]{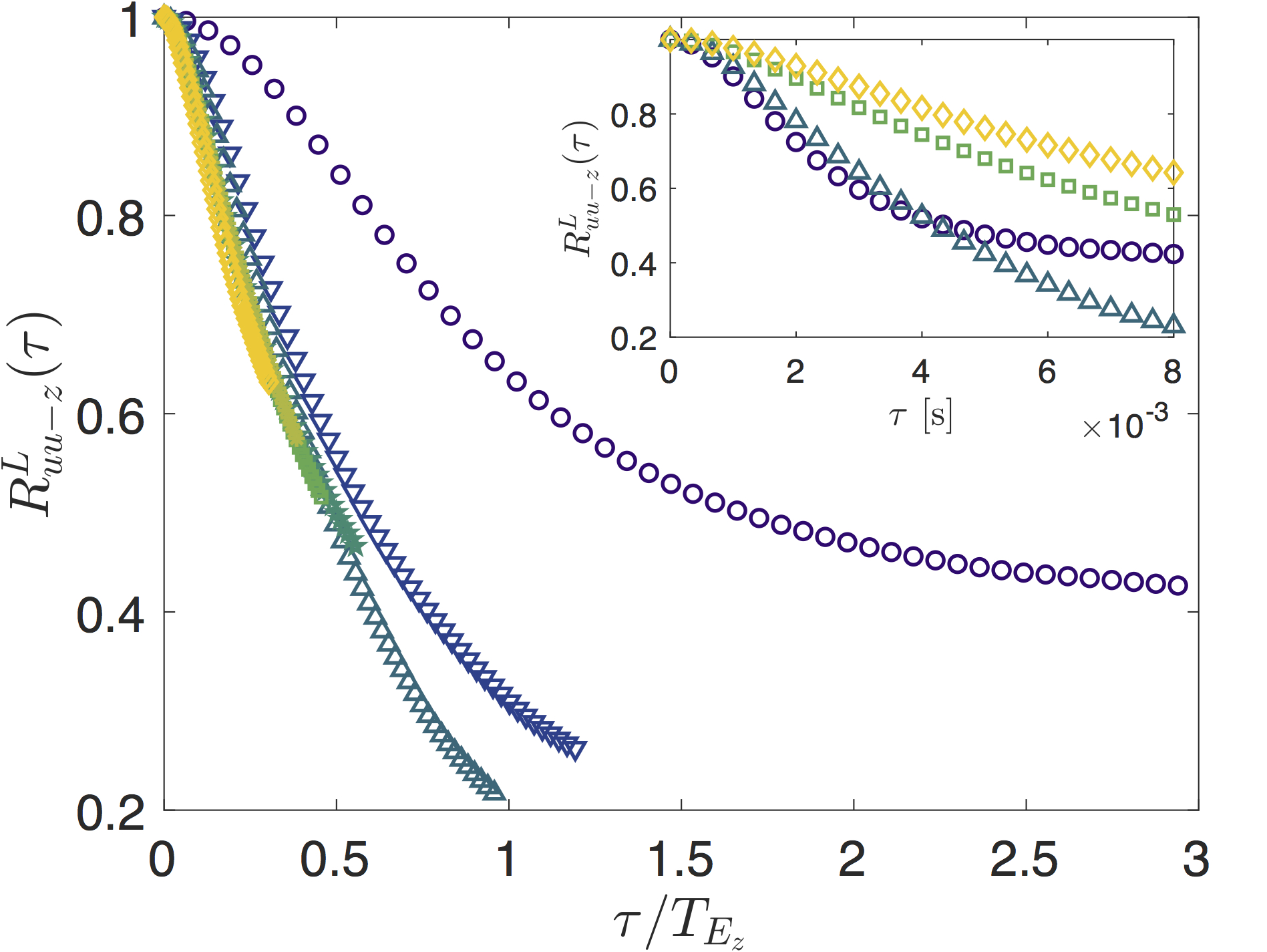}
\hspace{0.15cm}
\includegraphics[height=0.18\textheight]{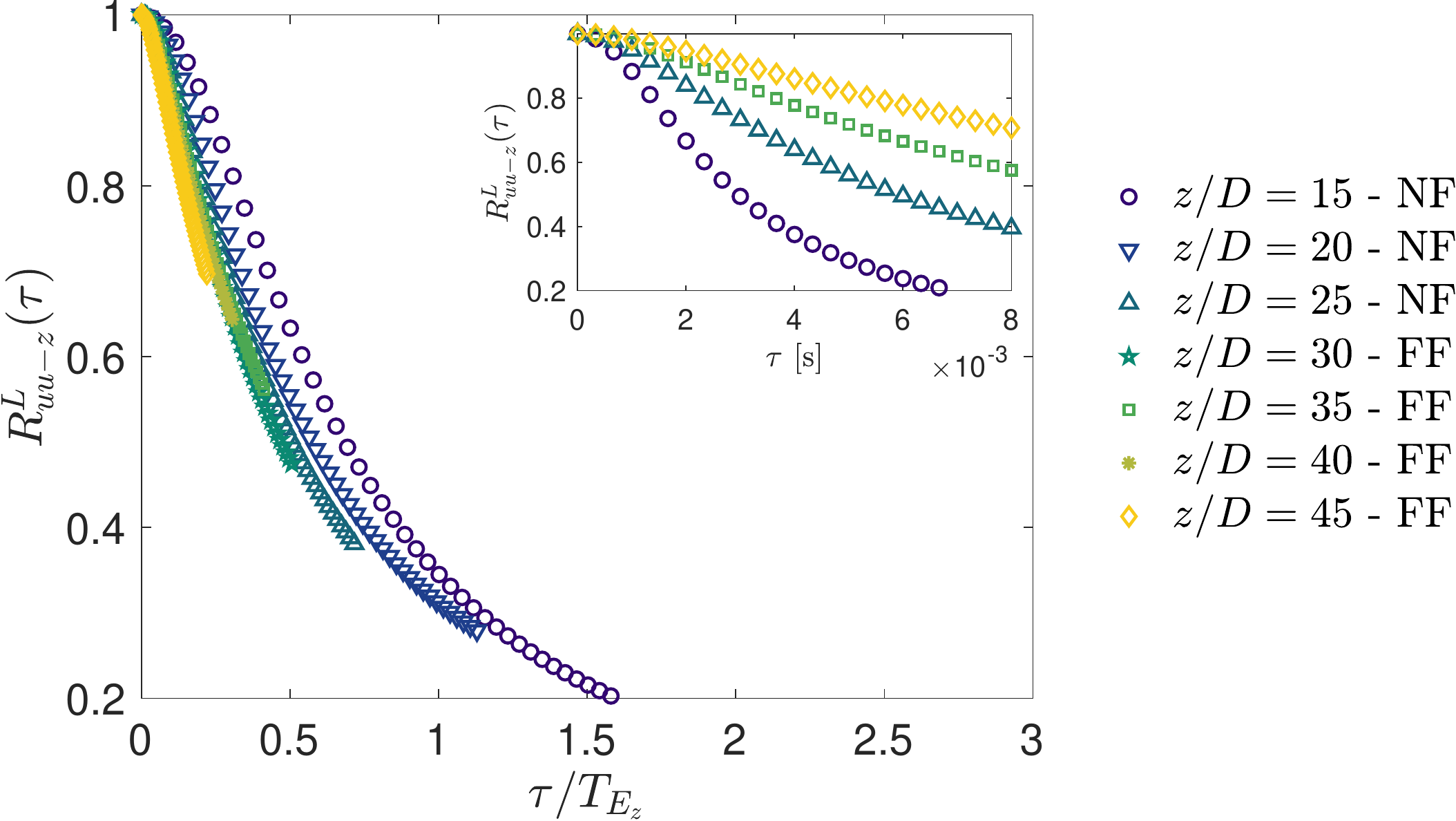}
\end{center}
\caption{Normalised Lagrangian correlation of the axial velocity for the compensated time lag $\tau/T_{E_z}$. Inset provides the Lagrangian correlation as a function of the dimensional time lag $\tau$ for the same seven downstream locations previously considered. Locations are (a)~along the centerline ($r=0$) and (b)~at the jet half-width ($r=r_{1/2}$) for all downstream positions.\label{fig:ruu_lag}}
\end{figure}
The two-point correlation functions of the Lagrangian axial velocity as a function of the compensated time lag, $\tau/T_{E_z}$, are presented in figure~\ref{fig:ruu_lag}(a) where $R^L_{uu-z}(\tau) = \langle u_z(t+\tau) u_z(t) \rangle$. It can be seen that, as for the structure functions previously discussed, the stationarisation results in a remarkable collapse of the correlation functions, in particular at $z/D > 20$. Note that the small scale discrepancy observed for $S^L_{2,z}$ is also expected to be present for the correlation function, which carries essentially the same information; it is however less emphasised due to the linear (rather than logarithmic) scale used to represent the correlation function. The observed agreement between the two-point correlation functions confirms again the Lagrangian self-similarity hypothesis at inertial scales, resulting in two-point correlation functions of the stationarised variables which do not depend on the downstream position of the particles as they evolve along the jet (beyond $z/D \gtrsim 20$). 

Although the shortness of the trajectories does not allow to directly explore the large scale, fully decorrelated, regime (where $R_{uu,i}$ vanishes), the observed collapse at intermediate scales allows speculation that the self-similarity hypothesis may also extend to the large scales. This would lead, in particular, to a univocal relation between the Lagrangian correlation time (defined as $T_{L_i} = \int_0^\infty R^L_{uu,i}(\tau) \:\mathrm{d}\tau$) and the Eulerian timescale at all positions along the jet (except in the very near-field, where Lagrangian two-point correlation clearly deviates). This point will be further tested in the subsection~\ref{subsec:lagParam} where we estimate $T_L$ based on appropriate fits (exponential or double exponential \citep{sawford1991reynolds}) of the Lagrangian two-point correlation, supporting the validity of self-similarity in the large scales and the univocal link between $T_L$ and $T_E$.

The subsection is concluded by providing, in figure~\ref{fig:ruu_lag}(b), a test of the Lagrangian self-similarity hypothesis when off-axis dynamics is considered. The original stationarisation proposed by \citet{batchelor1957diffusion} used centerline power-laws for a self-similar jet to compensate the Lagrangian velocity and time. As discussed in section~\ref{subsec:stat}, these formulas have been generalised (compatible with Batchelor's approach in the centerline), using actual local measurements of Eulerian properties rather than prescribed centerline power-laws. The stationarisation transformations can therefore be applied at any arbitrary position along particles trajectories. Figure~\ref{fig:ruu_lag}(b) explores the application of the proposed stationarisation considering trajectories passing through a ball centred off-axis, at a radial location of $r = r_{1/2}(z)$, instead of $r = 0$. As for the centerline analysis, the correlation functions of the stationarised variables collapse for all locations $z/D > 20$. This substantiates the generalised stationarisation technique, and its application to locations beyond the centerline. Although the present study focus on diffusion of particles near the jet centerline, this result motivates future dedicated studies to explore more deeply the generalised Lagrangian stationarisation for off-axis statistics as well as for other inhomogeneous flows (such as von K\'arm\'an flows, which are widely used for Lagrangian studies of turbulence).

\subsection{Evolution of Lagrangian parameters}\label{subsec:lagParam}
This subsection provides the estimates of $C_0$ and $T_L$, their streamwise evolution along the jet centerline, their connection to Eulerian properties of the jet and the reliability of Lagrangian stochastic models derived for HIST \citep{sawford1991reynolds} to address the stationarised Lagrangian dynamics of the jet. Investigations are made into these quantities for both axial and radial components of the velocity.

\begin{table}
\centering
\begin{tabular}{cccc|ccc}
$z/D$ & $C_{0_z}$ & $T_{L_z}$ $[\SI{}{ms}]$ & $T_{E_z}/T_{L_z}$ & $C_{0_r}$ & $T_{L_r}$ $[\SI{}{ms}]$ & $T_{E_{r\theta}}/T_{L_r}$\\[3pt]
15 & 1.4 & 4.5 & 0.6 & 1.9 & 1.4 & 2.1\\
25 & 2.7 & 5.3 & 1.6 & 3.2 & 2.3 & 2.3\\
35 & 3.2 & 11.1 & 1.5 & 3.0 & 5.3 & 2.5\\
45 & 3.0 & 15.9 & 1.8 & 2.8 & 8.9 & 2.6\\
\end{tabular}
\caption{Lagrangian parameters of the jet on the axis for different $z/D$ positions\label{tab:lag_parameters}.}
\end{table}
Table~\ref{tab:lag_parameters} presents these Lagrangian parameters of the jet for different $z$ locations in the near and far-fields. The scaling constant $C_{0_z}$ is observed in the compensated Lagrangian structure functions, figure~\ref{fig:s2_lag}(b). The Lagrangian integral time scale $T_{L_z}$ is estimated based on an exponential fit of the velocity correlation curves, due to the lack of experimental data for large time lags, figure~\ref{fig:ruu_lag}(a). Lagrangian correlation functions are indeed known (at least in HIST) to be well fitted by double exponential functions, and even simple exponential functions at sufficiently large Reynolds number, if the focus is on the estimate of inertial and large scales behaviour \citep{sawford1991reynolds}. In the present case, the fit by a simple exponential ($e^{-\tau/T_{L_z}}$) leads to very similar estimates of $T_{L_z}$ compared to a more sophisticated double exponential fit. Corresponding radial quantities are extracted in the same way by considering $S^L_{2-r}(\tau)$ and $R^L_{uu-r}(\tau)$.

\begin{figure}
\centerline{\includegraphics[width=0.85\textwidth]{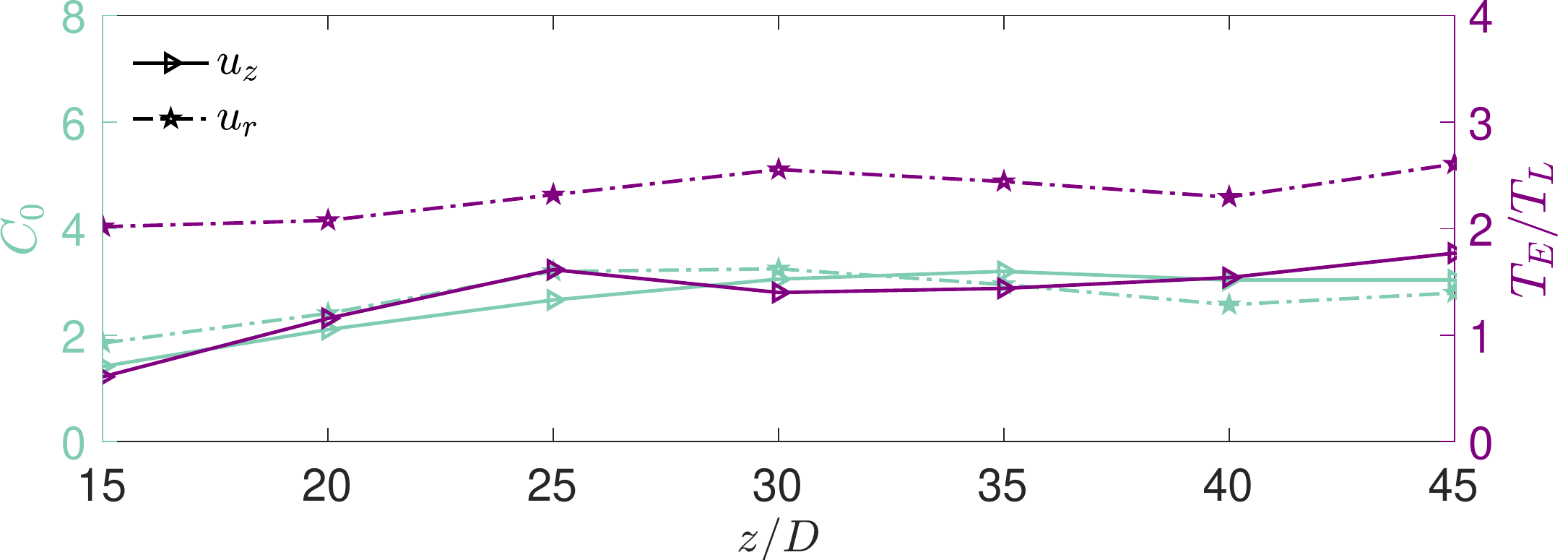}}
\caption{Evolution of the scaling constant $C_0$ (\emph{left}) and the ratio of the integral time scales $T_E/T_L$ (\emph{right}) as a function of downstream location within the jet centre. The axial ($-$) and radial (-$\cdot$-) components are both presented.\label{fig:c0}}
\end{figure}
$C_{0_z}$ is found to converge to a constant value $C_{0_z} \simeq 3.2$ at $z/D > 30$. This is more easily observed in figure~\ref{fig:c0} which provides the evolution of the Lagrangian parameters as a function of the downstream position $z$. The asymptotic far-field value of $C_{0_z}$ can be compared to values reported in the literature for $C_0$. A relationship presented by \citet{lien2002kolmogorov} accounting for finite Reynolds number effects on $C_0$ suggests an altered $C_0^*(\Rey_\lambda) = C^\infty_0 [1-(0.1\Rey_\lambda)^{-1/2}]$ where according to \citet{sawford1991reynolds}, $C^\infty_0 \simeq 7.0$. This gives an estimated $C_0^*$ of 5.6 for the Reynolds number corresponding to the present study as a benchmark value. As previously mentioned, discrepancies exist between numerous studies of this parameter, for example a $C_0^*$ of 4.8 was extracted for direct numerical simulation data with $\Rey_\lambda = 240$ by \citet{sawford2001lagrangian}, while experimental data taken between two counter-rotating disks at $\Rey_\lambda = 740$ produced a $C_0^*$ of 2.9 \citep{mordant2001measurement}. \citet{ouellette2006quantitative} found in a similar flow at $\Rey_\lambda\simeq 200$ an anisotropic behaviour, with $C_0^* \simeq 3.5$ for the velocity component aligned with the axis of rotation of the disks and $C_0^* \simeq 5.5$ for the transverse components. It is therefore difficult to be fully conclusive regarding the expected value of $C_0$ in our case, as it appears to be non-universal and not only dependent on the Reynolds number, but for a given Reynolds number to also depend on specific geometrical properties of the considered flow. It is observed however that the measured value of $C_{0_z}$ in the jet is in the same range of magnitude as other studies in different flows at similar Reynolds number. With regard to anisotropy, table~\ref{tab:lag_parameters} and figure~\ref{fig:c0} suggest that $C_{0_z}$ and $C_{0_r}$ behaves almost identically along the jet, $C_{0_z}$ converging to a value of 3.0 and $C_{0_r}$ to a value of 2.8. This indicates, on one hand that Lagrangian dynamics exhibit a level of isotropy, and on the other hand that at a specific location downstream, $C_0$ becomes independent of axial location and hence supports the idea that inertial Lagrangian statistics reaches self-similarity.

Regarding the Lagrangian correlation time scale both $T_{L_z}$ and $T_{L_r}$ increase with increasing axial distance, with $T_{L_z}$ being however significantly larger (about double) than $T_{L_r}$, see table~\ref{tab:lag_parameters}. Large scale Lagrangian dynamics therefore exhibits a persistent anisotropy, somehow more pronounced than the anisotropy previously reported for the Eulerian integral time scales (see for instance $T_{E_z}$ and $T_{E_{r\theta}}$ in table~\ref{tab:eul_parameters}). To further compare Lagrangian and Eulerian large scales properties, the ratio of the Eulerian to Lagrangian integral time scales, for both the axial and the radial components of velocity is provided in table~\ref{tab:lag_parameters} and figure~\ref{fig:c0}. For all locations, the Eulerian to Lagrangian time scale ratio for the radial component is notably larger (about double) that of the axial component. 
The axial component trends are consistent with similar results reported by \citet{gervais2007acoustic}, wherein $T_{E_z}/T_{L_z}$ was found to be less than one in the near-field, and to evolve towards a value greater than one (between 1.3 and 1.8) as the jet develops. Interestingly, in the well developed region, the Lagrangian dynamics decorrelates significantly faster compared to the Eulerian dynamics, as originally intuited by \citet{kraichnan1964relation}. This relation between Eulerian and Lagrangian time scales has been examined numerically by \citet{yeung2002lagrangian} where a ratio of $T_E/T_L = 1.28$ was found for HIST. This value is slightly lower than the value found in the present experiments, but is still consistent with a Lagrangian dynamics decorrelating faster than the Eulerian. 

Since the study by \citet{kraichnan1964relation} who suggested that $T_E/T_L > 1$, a similar prediction has been made by \citet{sawford1991reynolds} based on simple Lagrangian stochastic modelling. In this approach, Eulerian and Lagrangian time scales can be simply related to each other via the scaling constant $C_0$:
\begin{equation}
T_E/T_L = C_0/2.
\label{eq:c0eq}
\end{equation}
As observed in figure~\ref{fig:c0}, this relation is tested against the experimental results for the axial and radial velocity components. Note that the limits of the axis for $T_E/T_L$ on the right of the figure are half the limit of the axis for $C_0$ on the left of the figure, therefore if $T_E/T_L = C_0/2$ holds, the curves for $T_E/T_L$ and for $C_0$ shall superimpose). For the axial component, the agreement is almost perfect at all locations, including in the near-field. This is not observed for the radial component, while the two curves exhibits proportionality, the ratio of time scales is nearly equal to the scaling constant $C_0$ at all presented locations.

\section{Lagrangian acceleration statistical analysis}\label{sec:acc}
This section explores the statistics of the Lagrangian acceleration, to further elucidate small scales dynamics and its evolution along the jet. Particularly, the role of finite particle size effects (which evolve along the jet, and therefore may be to blame for preventing self-similarity to be recovered until the very far-field, as discussed in section~\ref{subsec:lagS2}) and the associated dimensionless constant $a_0$ (appearing in the Heisenberg-Yaglom relation \citep{monin1975statistical}) are addressed. At the same time, investigation into the connection between Eulerian and Lagrangian dissipative time scales can be carried out to further probe the applicability of stochastic models. All analysis is performed on trajectories that pass through a sphere of radius $r_{1/2}(z)/3$, as it was done for the Lagrangian velocity analysis. Only the axial component of velocity is considered for the acceleration discussion (radial component gives almost identical conclusions).

\subsection{Acceleration variance}
The variance of acceleration components is traditionally characterised by the scaling constant $a_0$ through the Heisenberg-Yaglom relation \citep{monin1975statistical}:
\begin{equation}
\langle a_z^2 \rangle = a_0 \nu^{-1/2} \varepsilon^{3/2},
\label{eq:a0}
\end{equation}
where $\nu$ is the fluid viscosity and $\varepsilon$ is the dissipation rate. The acceleration variance is taken directly from the trajectories with the $dt$-method implemented to find the noiseless values of $\langle a_z^2 \rangle$ \citep{machicoane2017estimating}. This is done for different $z$ positions along the jet. $a_0$ is then deduced at the different positions as $\langle a_z^2\rangle \nu^{1/2}\varepsilon_z^{-3/2}$, where $\varepsilon_z$ is the estimate of the dissipation rate at the considered position, based on the considered acceleration component.

\begin{figure}
\centerline{\includegraphics[width=0.55\textwidth]{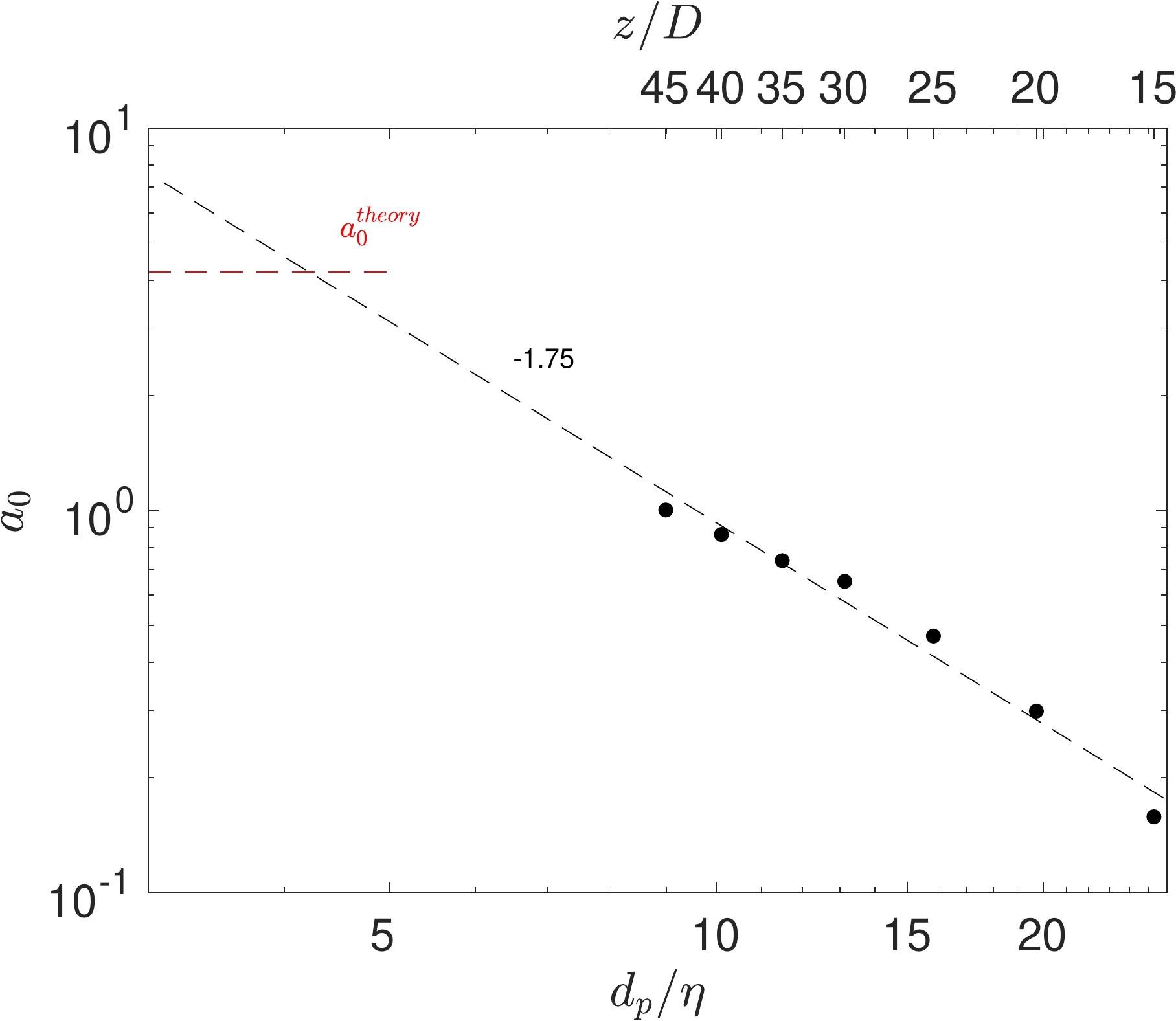}}
\caption{Evolution of the scaling constant of acceleration $a_0$ along the centerline as a function of the finite particle size $d_p/\eta$.\label{fig:a0}}
\end{figure}
Acceleration variance, and therefore the dimensionless constant $a_0$, is known to be highly sensitive to particle finite size effects and to converge to the value expected for actual tracers only when the normalised particle diameter $d_p/\eta \lesssim 5$ \citep{voth2002measurement, qureshi2007turbulent, calzavarini2009acceleration, volk2011dynamics}, where $\eta$ is the Kolmogorov length scale (see table~\ref{tab:eul_parameters}). In the present study the ratio $d_p/\eta$ varies typically between 9 and 25 depending on the distance to the nozzle. Therefore, the constant $a_0$, as a function of the normalised particle size $d_p/\eta$ (bottom axis) and of the downstream normalised location $z/D$ (top axis), is provided in figure~\ref{fig:a0}. Included is a power-law fit of -1.75 and a red dashed line of the expected value (from numerical simulations of HIST), $a_0^{theory} \simeq 4.2$, calculated from \citet{sawford1991reynolds}. The power-law of -1.75 provides the expected $a_0^{tracer}$ value of a true tracer through extrapolating the trend as $d_p/\eta \rightarrow 5$, from which a value of $a_0^{tracer} \simeq 3.0$ is found, in reasonable agreement with values reported in previous experimental studies in von K\'arm\'an flows \citep{voth2002measurement} and numerical simulations in HIST \citep{sawford1991reynolds, vedula1999similarity} for similar Reynolds number. Furthermore, the power-law fit intersects with the theoretical value of $a_0^{theory}$ at $d_p/\eta \lesssim 5$, what is generally considered as the diameter for which finite size effects become noticeable. These observations suggest that acceleration statistics in the jet should eventually behave for tracers as in HIST, without a major influence of large scale inhomogeneity of the jet. With the present considered particles (with $d_p \simeq \SI{250}{\micro m}$) the tracer behaviour is expected to be reached at a downstream distance $z/D \simeq 65$, which is out of reach of the present data set. To deepen this question, it would be interesting to perform further experiments specifically dedicated to acceleration measurements, by considering either smaller particles or further downstream distances.

Regarding finite size effects, previous studies have reported in HIST a power-law dependency of $a_0$ on particle size, with $a_0 \propto (d_p/\eta)^{-2/3}$ \citep{qureshi2007turbulent, brown2009acceleration}, while a study by \citet{volk2011dynamics} of von K\'arm\'an dynamics report a slightly steeper decay with an exponent -0.81. In the present study, an even steeper decrease of constant $a_0$ is observed with particle size, with an exponent -1.75, about double of the values reported previously. This stronger dependence of $a_0$ on particle size remains to be elucidated. It is likely due to a coupling between the finite size effects and the streamwise dependence of turbulent properties in the jet, although further investigation would be necessary to further explore this point.

\subsection{Acceleration two-point correlation}
Beyond the value of $a_0$, acceleration statistics are also of great interest as they reflect the Lagrangian dissipative dynamics of the particles. In particular, they give access to the dissipative timescale of the Lagrangian dynamics, traditionally defined based on $\tau_0$, the zero-crossing time of the acceleration two-point correlation function, $R_{aa,z}(\tau) = \langle a_z(t+\tau) a_z(t) \rangle$ with $R_{aa,z}(\tau_0) = 0$. Acceleration two-point correlation is estimated with a variant of the $dt$-method \citep{machicoane2017estimating}. Briefly, the acceleration two-point correlation is obtained from second order position increments $d^2x$ according to the relation
\begin{equation}
R_{d^2xd^2x}(\tau,dt) = R_{\hat{a}\hat{a}}(\tau) dt^4
+R_{d^2bd^2b}(\tau,dt) + \mathcal{O}(dt^6),
\end{equation}
where $R_{d^2bd^2b}$ represents the contribution of noise.
A polynomial fit of $dt^4$ is implemented to extract the true correlation values of $R_{\hat{a}\hat{a}}$, eliminating the noise contribution. This method is extended, herein, to more accurately describe the correlation of acceleration of the given data set.

\begin{figure}
\centerline{\includegraphics[width=0.6\textwidth]{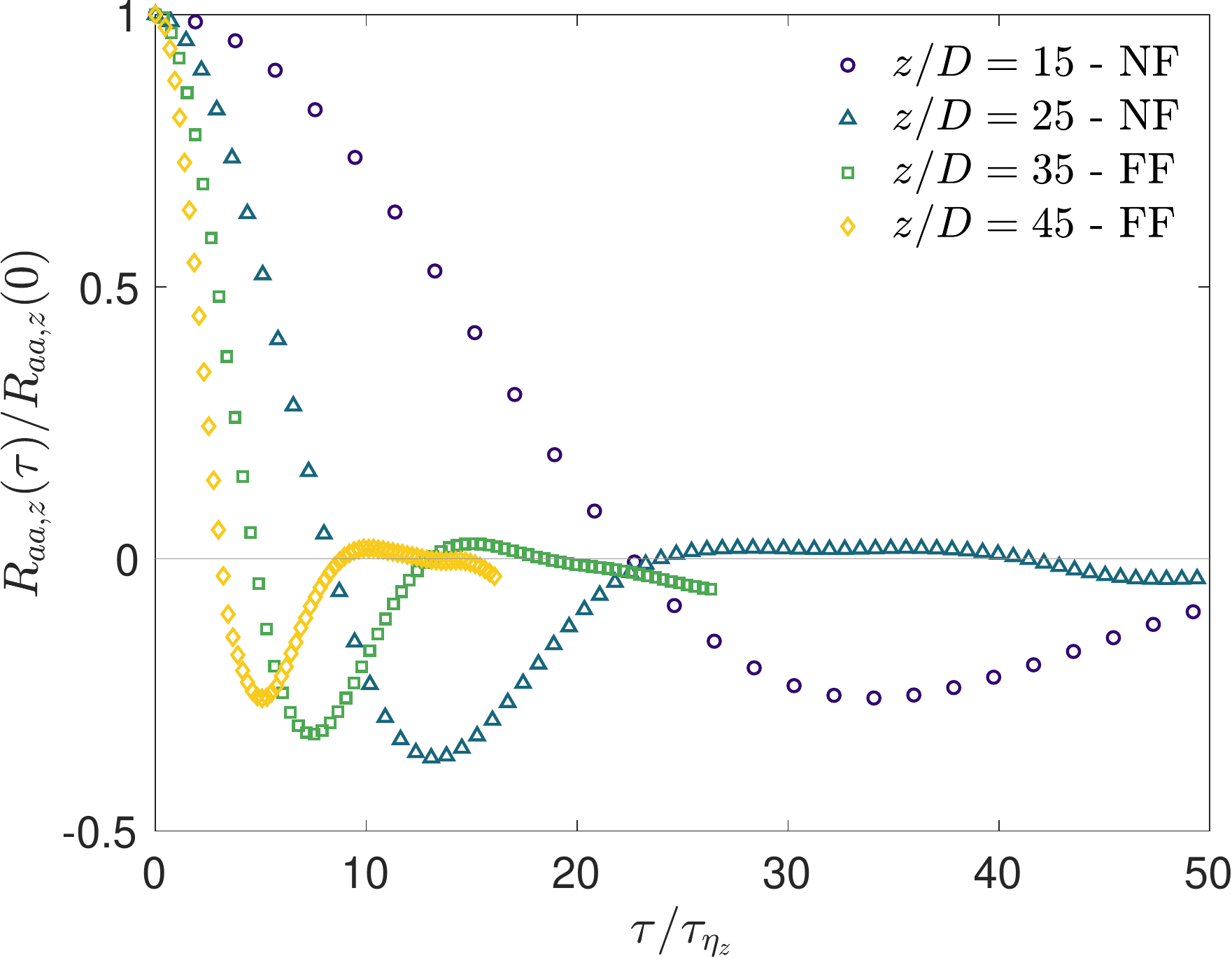}}
\caption{Normalised axial acceleration correlation on the axis as a function of time lag normalised by the Kolmogorov time scale.\label{fig:Raa}}
\end{figure}
\begin{figure}
\centerline{\includegraphics[width=0.85\textwidth]{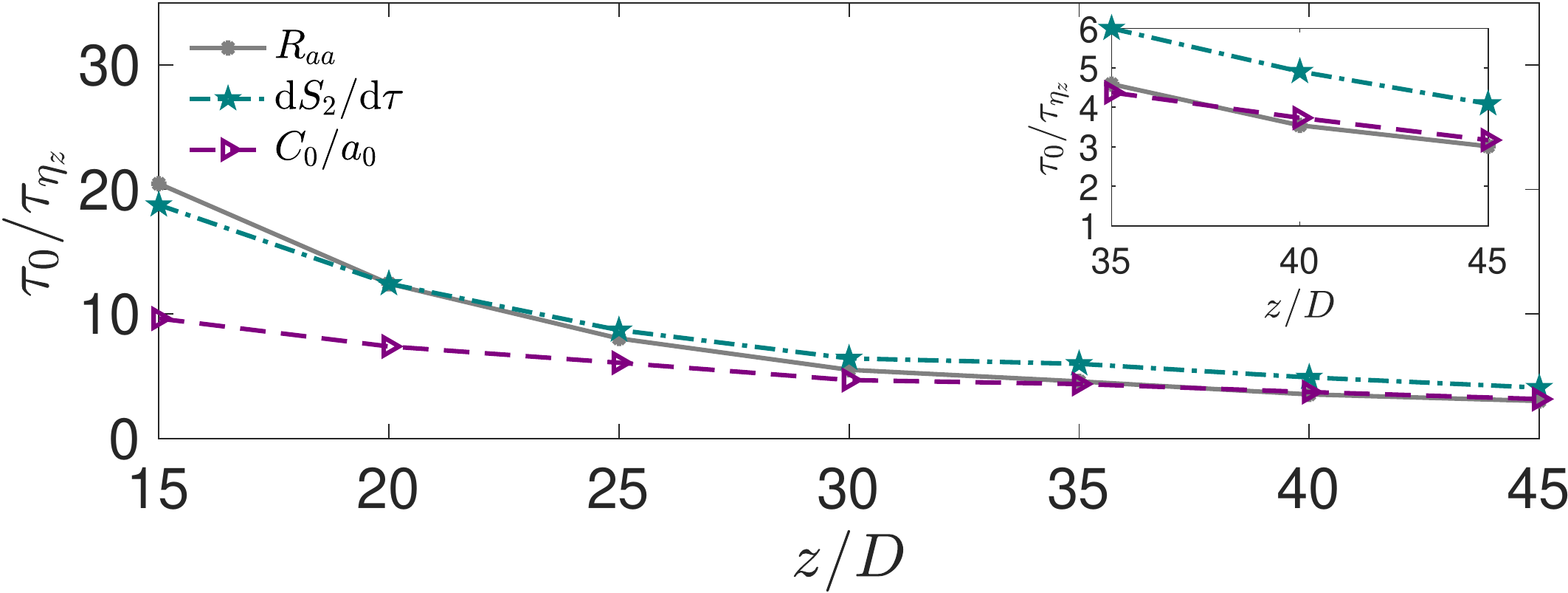}}
\caption{The zero-crossing of the acceleration correlation normalised by the Kolmogorov time scale as a function of the downstream location along the centre of the jet. Three estimations are presented based on the acceleration correlation $R_{aa}$, the derived second-order structure function $\mathrm{d}S^L_2/\mathrm{d}\tau$, and the model driven values obtained from $C_0/a_0$.\label{fig:tau_0}}
\end{figure}

The two-point correlation of acceleration is presented in figure~\ref{fig:Raa} for four downstream locations along the centerline, where the time lag has been normalised by $\tau_\eta$. It has been noted in previous studies that for tracers $\tau_0 \simeq 2.2\tau_\eta$ \citep{yeung1989lagrangian, calzavarini2009acceleration, volk2008acceleration}. For the current study, the the zero-crossing time is not unequivically close to $\tau_{\eta}$ and therefore the ratio $\tau_0/\tau_\eta$ depends on the location of the measurement. The expected value of 2 is only approached in the farthest downstream locations within the jet. The solid line in figure~\ref{fig:tau_0} shows the downstream evolution of the zero-crossing time $\tau_0$. As for $a_0$ the observed streamwise dependency of $\tau_0/\tau_\eta$ is likely due to finite size effects, which have been reported in HIST to be affected by finite size effects \citep{volk2008acceleration, calzavarini2009acceleration}. It shall be noted though, that $\tau_0/\tau_\eta$ seems to eventually approach the expected value of nearly 2 for the farthest positions (and hence for the smallest $d_p/\eta$ ratios), presented in the inset of figure~\ref{fig:tau_0}. Following the considerations previously discussed on the trends of $a_0$, it could then be expected that the actual tracer behaviour (free of finite size effects) would be fully recovered for $\tau_0$ near $z/D\simeq 65$, with a ratio $\tau_0/\tau_\eta$ of the same order of what is usually reported for HIST.

Beyond the discussed finite size effects, acceleration correlation is also insightful to shed further light on the Lagrangian properties of the jet of relevance for the application to diffusion problems, as motivated in the introduction.

First, the stationarisation of velocity \emph{\`a la} Batchelor can be tested further tested by recalling that for any random stationary signal $\xi$, the two-point correlation of the derivative of $\xi$, $R_{\dot{\xi}\dot{\xi}}$, is simply related to the second derivative of the two-point correlation of $\xi$: $R_{\dot{\xi}\dot{\xi}} = -\ddot{R}_{\xi\xi}$ (derivatives are denoted in dot notation). In the present case, this relation gives that the zero-crossing of acceleration correlation corresponds to an inflection point of the velocity two-point correlation. If Lagrangian stationarity holds, $\tau_0$ can therefore be simply extracted from the peak of the derivative of the second-order structure function, $\mathrm{d}S^L_2/\mathrm{d}\tau$ (figure not included). The corresponding values are presented in figure~\ref{fig:tau_0} (dot-dashed line) which exhibit a fair agreement with the direct estimate of $\tau_0$ from $R_{aa}$. This observation supports the validity of the proposed stationarisation procedure at each explored location independently. Although, finite size effects influence the streamwise dependence of $\tau_0$ therefore impeding the validation of the small scale Lagrangian self-similarity based on streamwise evolution of $\tau_0$ (or $a_0$).

Second, the relevance of stochastic models to characterise the Lagrangian dynamics (and therefore to predict diffusion properties) can be further tested from the acceleration timescales. As presented in section~\ref{subsec:lagParam}, simple (Langevin) stochastic models accurately predict large scale properties, such as the connection between Lagrangian and Eulerian integral timescales and $C_0$. Two-time stochastic models \citep{sawford1991reynolds} also predict a similar relation for the small Eulerian and Lagrangian timescales, involving the constant $C_0$ and $a_0$ (see \citet{huck2019lagrangian}). Namely, the prediction from such models can be written as
\begin{equation}
\tau_a = \int_0^{\tau_0} R_{aa}(\tau) \:\mathrm{d}\tau = \frac{C_0}{2a_0}\tau_\eta.
\label{eq:tau_int}
\end{equation}
Neglecting the curvature of $R_{aa}$ at the origin, the integral $\int_0^{\tau_0} R_{aa}(\tau) \:\mathrm{d}\tau$ can be approximated as $\tau_a \simeq \thalf \tau_0$ (because of the curvature, it is actually slightly larger than that). It is therefore expected from stochastic models that $\tau_0/\tau_\eta \simeq C_0/a_0$. The dashed line in figure~\ref{fig:tau_0} represents the downstream evolution of the ratio $C_{0_z}/a_{0_z}$ extracted from the measurements. Neglecting the near-field locations of $z/D < 25$, it can be seen that the agreement is also adequate when compared to the two previously presented independent estimations of $\tau_0/\tau_\eta$.

\section{Conclusion}\label{sec:conc}
Particle tracking velocimetry was implemented to create three component jet trajectories in three-dimensional space. Generation of such a large scale database facilitates the study of how fundamental Lagrangian parameters behave when exposed to a highly anisotropic and inhomogeneous flow field. The Lagrangian self-similarity theory of turbulent diffusion by \citet{batchelor1957diffusion} has been applied to account for the Lagrangian instationarity of the flow field due to the spatial Eulerian inhomogeneity. The stationarisation technique leading to Lagrangian self-similarity is validated in the far-field of the jet for Lagrangian inertial scales dynamics by the collapse of the Lagrangian velocity structure functions and correlation profiles (after a given location downstream) for the stationarised variables. The Lagrangian self-similarity is also validated for the large scales, as the Lagrangian and Eulerian time scales are found to be univocally tight in the far-field of the jet. For the small scales Lagrangian dynamics, self-similarity is only observed in the farthest downstream locations explored. This is attributed to the impact of particle finite size effects which evolve along the jet axis and therefore influence the small scale Lagrangian dynamics differently depending on the downstream position, as confirmed by the acceleration statistics. Further studies, with experiments specifically dedicated to small scale (acceleration) measurements of small tracers would be required to draw final conclusions concerning the small scale Lagrangian self-similarity. In turn, this confirmation of the validity of the Lagrangian self-similarity at inertial and large scales is an important element supporting Batchelor's extension of Taylor's stationary theory of turbulent diffusion to the case of self-similar jets and wakes where particles have a non-stationary Lagrangian dynamics.

Regarding the inertial scales of the Lagrangian dynamics, results indicate that the Lagrangian scaling constant, $C_0$, is a function of downstream location in the near-field and eventually converges (around $z/D = 30$) to a value of the order of 3, with a small ($\sim 10\%$) difference between axial and radial components, indicating a weak role of anisotropy on inertial scale Lagrangian dynamics in the jet. It is noted that this value may be Reynolds number dependent (its order of magnitude is consistent though with HIST simulations and experiments carried in other flows at similar Reynolds number), and further studies in a jet configuration at different Reynolds number will be required to explore this dependency. 

The evolution of the Eulerian to Lagrangian integral time scale ratio shows convergence towards $T_E/T_L \simeq 1.8$ around $z/D = 25$ for the axial velocity timescales and $T_E/T_L \simeq 2.6$ by the same location downstream for the radial based timescale ratio. This points towards three interesting observations: (i)~In the well developed region of the jet, the Lagrangian dynamics decorrelates faster (about twice faster) than the Eulerian (as predicted for HIST by \citet{kraichnan1964relation}); (ii)~The ratio between Lagrangian and Eulerian integral scales is about 40\% larger for the radial component compared to the axial, what is to be related to the large scale anisotropy of the jet; (iii)~In spite of this difference, sufficient agreement is found between the measured ratio for these time scales and the prediction from simple stochastic models for HIST, $T_E/T_L \simeq C_0/2$ (the agreement is favourable between the axial based parameters while the value predicted by the model underestimate the actual time scale ratio for in the radial direction).

Considering the small scale dynamics, the normalised acceleration variance shows a strong dependence on the downstream location from the nozzle, presumably associated to finite particle size effects, which are known to influence acceleration when $d_p/\eta > 5$ typically. This presumably explains why self-similarity is not fully recovered at small scales in the present study, as tracer like behaviour for acceleration would only be recovered around $z/D\simeq 65$. Besides, the power-law slope of $a_0$ as a function of $d_p/\eta$ found in the current study is larger than in previous studies in HIST and von K\'arm\'an flows, suggesting that the jet dynamics interplay with finite size effects. The zero-crossing of the acceleration correlation also demonstrates a strong dependence on the downstream location from the nozzle, converging towards typical values ($\tau_0/\tau_\eta\simeq 2$) only at the farthest position explored ($z/D\gtrsim 40$). Although the actual value of $\tau_0$ is likely also altered due to finite size effects, the agreement between several independent estimates of $\tau_0$ supports on the one hand the validity of the proposed stationarisation method and on the other hand the relevance of simple stochastic approaches to link (in the far field) the Eulerian and Lagrangian dissipative time scales to the experimentally determined constants $C_0$ and $a_0$.

The ability of the implemented stationarisation technique provides adequate methods for calculating the scaling constant, a non-trivial task within an inhomogeneous flow field. Overall, after a proper stationarisation, the Lagrangian properties for the jet are interestingly found to match reasonably well the behaviours previously reported for HIST. From the perspective of building simple and practical diffusion models, the success of the method validates Batchelor's extension of Taylor's theory, providing estimates of turbulent diffusion properties based on the Lagrangian second-order structure function (or two-point correlation function) of velocity. Further, the relations presented between the Eulerian and Lagrangian time scales (both integral and dissipative) suggests that simple stochastic modelling is well suited to find reasonable estimates of such correlation functions. Actually, based on these models, the simple knowledge of the constants $a_0$ and $C_0$ may be sufficient to build reasonable proxies (with exponential or double exponential functions) of these correlations to be used for estimating turbulent diffusion properties.

\section*{Acknowledgements}
B.V., S.S. and R.B.C. are supported by U.S. National Science Foundation grant (GEO1756259). R.B.C. is also thankful for the support provided through the Fulbright Scholar Program. B.V., L.C., R.V. and M.B. benefit from the financial support of the Project IDEXLYON of the University of Lyon in the framework of the French program Programme Investissements d’Avenir (ANR-16-IDEX-0005). L.C. is supported by ANR grants Liouville ANR-15-CE40-0013 and by the Simons Foundation Award ID: 651475.

\bibliographystyle{jfm}
\bibliography{Jetlagbib.bib}

\begin{thebibliography}{47}
\expandafter\ifx\csname natexlab\endcsname\relax\def\natexlab#1{#1}\fi
\def\au#1{#1} \def\ed#1{#1} \def\yr#1{#1}\def\at#1{#1}\def\jt#1{\textit{#1}}
  \def\bt#1{#1}\def\bvol#1{\textbf{#1}} \def\vol#1{#1} \def\pg#1{#1}
  \def\publ#1{#1}\def\arxiv#1{#1}\def\org#1{#1}\def\st#1{\textit{#1}}

\bibitem[Batchelor(1957)]{batchelor1957diffusion}
{\sc \au{Batchelor, G.~K.}} \yr{1957}  \at{Diffusion in free turbulent shear
  flows}.  \jt{J.~Fluid Mech.}  \bvol{3}~(1),  \pg{67--80}.

\bibitem[Berk \& Coletti(2020)]{berk2020transport}
{\sc \au{Berk, T.} \& \au{Coletti, F.}} \yr{2020}  \at{Transport of inertial
  particles in high-reynolds-number turbulent boundary layers}.  \jt{Journal of
  Fluid Mechanics}  \bvol{903}.

\bibitem[Bourgoin \& Huisman(2020)]{bourgoin2020using}
{\sc \au{Bourgoin, M.} \& \au{Huisman, S.~G.}} \yr{2020}  \at{Using
  ray-traversal for {3D} particle matching in the context of particle tracking
  velocimetry in fluid mechanics}.  \jt{Rev. Sci. Instrum.}  \bvol{91}~(8),
  \pg{085105}.

\bibitem[Brown {\em et~al.\/}(2009)Brown, Warhaft \&
  Voth]{brown2009acceleration}
{\sc \au{Brown, R.~D.}, \au{Warhaft, Z.} \& \au{Voth, G.~A.}} \yr{2009}
  \at{Acceleration statistics of neutrally buoyant spherical particles in
  intense turbulence}.  \jt{Phys. Rev. Lett.}  \bvol{103}~(19),  \pg{194501}.

\bibitem[Burattini {\em et~al.\/}(2005)Burattini, Antonia \&
  Danaila]{burattini2005similarity}
{\sc \au{Burattini, P.}, \au{Antonia, R.~A.} \& \au{Danaila, L.}} \yr{2005}
  \at{Similarity in the far field of a turbulent round jet}.  \jt{Phys. Fluids}
   \bvol{17}~(2),  \pg{025101}.

\bibitem[Calzavarini {\em et~al.\/}(2009)Calzavarini, Volk, Bourgoin,
  L\'ev\^eque, Pinton \& Toschi]{calzavarini2009acceleration}
{\sc \au{Calzavarini, E.}, \au{Volk, R.}, \au{Bourgoin, M.}, \au{L\'ev\^eque,
  E.}, \au{Pinton, J.-F.} \& \au{Toschi, F.}} \yr{2009}  \at{Acceleration
  statistics of finite-sized particles in turbulent flow: the role of {Fax\'en}
  forces}.  \jt{J.~Fluid Mech.}  \bvol{630},  \pg{179--189}.

\bibitem[Cermak(1963)]{cermak1963lagrangian}
{\sc \au{Cermak, J.~E.}} \yr{1963}  \at{Lagrangian similarity hypothesis
  applied to diffusion in turbulent shear flow}.  \jt{J.~Fluid Mech.}
  \bvol{15}~(1),  \pg{49--64}.

\bibitem[Corrsin(1943)]{corrsin1943investigation}
{\sc \au{Corrsin, S.}} \yr{1943} Investigation of flow in an axially
  symmetrical heated jet of air. Nat. Adv. Comm. f.~Aeron., Adv. Conf. Rep.
  3L23, Wartime Report W-94.

\bibitem[Gervais {\em et~al.\/}(2007)Gervais, Baudet \&
  Gagne]{gervais2007acoustic}
{\sc \au{Gervais, P.}, \au{Baudet, C.} \& \au{Gagne, Y.}} \yr{2007}
  \at{Acoustic {Lagrangian} velocity measurement in a turbulent air jet}.
  \jt{Exp. Fluids}  \bvol{42}~(3),  \pg{371--384}.

\bibitem[Guezennec {\em et~al.\/}(1994)Guezennec, Brodkey, Trigui \&
  Kent]{guezennec1994algorithms}
{\sc \au{Guezennec, Y.~G.}, \au{Brodkey, R.~S.}, \au{Trigui, N.} \& \au{Kent,
  J.~C.}} \yr{1994}  \at{Algorithms for fully automated three-dimensional
  particle tracking velocimetry}.  \jt{Exp. Fluids}  \bvol{17}~(4),
  \pg{209--219}.

\bibitem[Hinze \& {Van Der Hegge Zijnen}(1949)]{hinze1949transfer}
{\sc \au{Hinze, J.~O.} \& \au{{Van Der Hegge Zijnen}, B.~G.}} \yr{1949}
  \at{Transfer of heat and matter in the turbulent mixing zone of an axially
  symmetrical jet}.  \jt{Flow Turb. Combust.}  \bvol{1},  \pg{435--461}.

\bibitem[Holzner {\em et~al.\/}(2008)Holzner, Liberzon, Nikitin, L\"uthi,
  Kinzelbach \& Tsinober]{holzner2008lagrangian}
{\sc \au{Holzner, M.}, \au{Liberzon, A.}, \au{Nikitin, N.}, \au{L\"uthi, B.},
  \au{Kinzelbach, W.} \& \au{Tsinober, A.}} \yr{2008}  \at{A {Lagrangian}
  investigation of the small-scale features of turbulent entrainment through
  particle tracking and direct numerical simulation}.  \jt{J.~Fluid Mech.}
  \bvol{598},  \pg{465--475}.

\bibitem[Huck {\em et~al.\/}(2019)Huck, Machicoane \& Volk]{huck2019lagrangian}
{\sc \au{Huck, P.~D.}, \au{Machicoane, N.} \& \au{Volk, R.}} \yr{2019}
  \at{Lagrangian acceleration timescales in anisotropic turbulence}.  \jt{Phys.
  Rev. Fluids}  \bvol{4}~(6),  \pg{064606}.

\bibitem[Hussein {\em et~al.\/}(1994)Hussein, Capp \&
  George]{hussein1994velocity}
{\sc \au{Hussein, H.~J.}, \au{Capp, S.~P.} \& \au{George, W.~K.}} \yr{1994}
  \at{Velocity measurements in a {high-Reynolds-number}, momentum-conserving,
  axisymmetric, turbulent jet}.  \jt{J.~Fluid Mech.}  \bvol{258},  \pg{31--75}.

\bibitem[Kennedy \& Moody(1998)]{kennedy1998particle}
{\sc \au{Kennedy, I.~M.} \& \au{Moody, M.~H.}} \yr{1998}  \at{Particle
  dispersion in a turbulent round jet}.  \jt{Exp. Therm. Fluid Sci.}
  \bvol{18}~(1),  \pg{11--26}.

\bibitem[Kim {\em et~al.\/}(2017)Kim, Liberzon \&
  Chamorro]{kim2017characterisation}
{\sc \au{Kim, J.-T.}, \au{Liberzon, A.} \& \au{Chamorro, L.~P.}} \yr{2017}
  \at{Characterisation of the eulerian and lagrangian accelerations in the
  intermediate field of turbulent circular jets}.  \jt{Journal of Turbulence}
  \bvol{18}~(1),  \pg{87--102}.

\bibitem[Kolmogorov(1941)]{kolmogorov1941local}
{\sc \au{Kolmogorov, A.~N.}} \yr{1941}  \at{The local structure of turbulence
  in incompressible viscous fluid for very large {Reynolds} numbers}.
  \jt{Dokl. Akad. Nauk SSSR}  \bvol{30}~(4),  \pg{301--305}.

\bibitem[Kraichnan(1964)]{kraichnan1964relation}
{\sc \au{Kraichnan, R.~H.}} \yr{1964}  \at{Relation between {Lagrangian} and
  {Eulerian} correlation times of a turbulent velocity field}.  \jt{Phys.
  Fluids}  \bvol{7}~(1),  \pg{142--143}.

\bibitem[Lien \& D’Asaro(2002)]{lien2002kolmogorov}
{\sc \au{Lien, R.-C.} \& \au{D’Asaro, E.~A.}} \yr{2002}  \at{The {Kolmogorov}
  constant for the {Lagrangian} velocity spectrum and structure function}.
  \jt{Phys. Fluids}  \bvol{14}~(12),  \pg{4456--4459}.

\bibitem[Lien {\em et~al.\/}(1998)Lien, D’Asaro \&
  Dairiki]{lien1998lagrangian}
{\sc \au{Lien, R.-C.}, \au{D’Asaro, E.~A.} \& \au{Dairiki, G.~T.}} \yr{1998}
  \at{Lagrangian frequency spectra of vertical velocity and vorticity in
  {high-Reynolds-number} oceanic turbulence}.  \jt{J.~Fluid Mech.}  \bvol{362},
   \pg{177--198}.

\bibitem[Lohse \& M\"uller-Groeling(1995)]{lohse1995bottleneck}
{\sc \au{Lohse, D.} \& \au{M\"uller-Groeling, A.}} \yr{1995}  \at{Bottleneck
  effects in turbulence: scaling phenomena in $r$ versus $p$ space}.  \jt{Phys.
  Rev. Lett.}  \bvol{74}~(10),  \pg{1747--1750}.

\bibitem[Machicoane {\em et~al.\/}(2019)Machicoane, Aliseda, Volk \&
  Bourgoin]{machicoane2019simplified}
{\sc \au{Machicoane, N.}, \au{Aliseda, A.}, \au{Volk, R.} \& \au{Bourgoin, M.}}
  \yr{2019}  \at{A simplified and versatile calibration method for multi-camera
  optical systems in {3D} particle imaging}.  \jt{Rev. Sci. Instrum.}
  \bvol{90}~(3),  \pg{035112}.

\bibitem[Machicoane {\em et~al.\/}(2017{\natexlab{{\em a\/}}})Machicoane, Huck
  \& Volk]{machicoane2017estimating}
{\sc \au{Machicoane, N.}, \au{Huck, P.~D.} \& \au{Volk, R.}}
  \yr{2017{\natexlab{{\em a\/}}}}  \at{Estimating two-point statistics from
  derivatives of a signal containing noise: Application to auto-correlation
  functions of turbulent {Lagrangian} tracks}.  \jt{Rev. Sci. Instrum.}
  \bvol{88}~(6),  \pg{065113}.

\bibitem[Machicoane {\em et~al.\/}(2017{\natexlab{{\em b\/}}})Machicoane,
  L\'opez-Caballero, Bourgoin, Aliseda \& Volk]{machicoane2017multi}
{\sc \au{Machicoane, N.}, \au{L\'opez-Caballero, M.}, \au{Bourgoin, M.},
  \au{Aliseda, A.} \& \au{Volk, R.}} \yr{2017{\natexlab{{\em b\/}}}}  \at{A
  multi-time-step noise reduction method for measuring velocity statistics from
  particle tracking velocimetry}.  \jt{Meas. Sci. Technol.}  \bvol{28}~(10),
  \pg{107002}.

\bibitem[Monin \& Yaglom(1975)]{monin1975statistical}
{\sc \au{Monin, A.~S.} \& \au{Yaglom, A.~M.}} \yr{1975} {\em Statistical Fluid
  Mechanics: Mechanics of Turbulence, Volume 2\/}.  \publ{MIT Press}.

\bibitem[Mordant {\em et~al.\/}(2004{\natexlab{{\em a\/}}})Mordant, L\'ev\^eque
  \& Pinton]{mordant2004experimental1}
{\sc \au{Mordant, M.}, \au{L\'ev\^eque, E.} \& \au{Pinton, J.-F.}}
  \yr{2004{\natexlab{{\em a\/}}}}  \at{Experimental and numerical study of the
  {Lagrangian} dynamics of high {Reynolds} turbulence}.  \jt{New J.~Phys}
  \bvol{6}~(1),  \pg{116}.

\bibitem[Mordant {\em et~al.\/}(2004{\natexlab{{\em b\/}}})Mordant, Crawford \&
  Bodenschatz]{mordant2004experimental2}
{\sc \au{Mordant, N.}, \au{Crawford, A.~M.} \& \au{Bodenschatz, E.}}
  \yr{2004{\natexlab{{\em b\/}}}}  \at{Experimental {Lagrangian} acceleration
  probability density function measurement}.  \jt{Physica~D}  \bvol{193}~(1-4),
   \pg{245--251}.

\bibitem[Mordant {\em et~al.\/}(2001)Mordant, Metz, Michel \&
  Pinton]{mordant2001measurement}
{\sc \au{Mordant, N.}, \au{Metz, P.}, \au{Michel, O.} \& \au{Pinton, J.-F.}}
  \yr{2001}  \at{Measurement of {Lagrangian} velocity in fully developed
  turbulence}.  \jt{Phys. Rev. Lett.}  \bvol{87}~(21),  \pg{214501}.

\bibitem[Ouellette {\em et~al.\/}(2006{\natexlab{{\em a\/}}})Ouellette, Xu \&
  Bodenschatz]{ouellette2006quantitative}
{\sc \au{Ouellette, N.~T.}, \au{Xu, H.} \& \au{Bodenschatz, E.}}
  \yr{2006{\natexlab{{\em a\/}}}}  \at{A quantitative study of
  three-dimensional {Lagrangian} particle tracking algorithms}.  \jt{Exp.
  Fluids}  \bvol{40}~(2),  \pg{301--313}.

\bibitem[Ouellette {\em et~al.\/}(2006{\natexlab{{\em b\/}}})Ouellette, Xu,
  Bourgoin \& Bodenschatz]{ouellette2006small}
{\sc \au{Ouellette, N.~T.}, \au{Xu, H.}, \au{Bourgoin, M.} \& \au{Bodenschatz,
  E.}} \yr{2006{\natexlab{{\em b\/}}}}  \at{Small-scale anisotropy in
  {Lagrangian} turbulence}.  \jt{New J.~Phys}  \bvol{8}~(6),  \pg{102}.

\bibitem[Pope(2000)]{pope2000turbulent}
{\sc \au{Pope, S.~B.}} \yr{2000} {\em Turbulent Flows\/}.  \publ{Cambridge
  University Press}.

\bibitem[Qureshi {\em et~al.\/}(2007)Qureshi, Bourgoin, Baudet, Cartellier \&
  Gagne]{qureshi2007turbulent}
{\sc \au{Qureshi, N.~M.}, \au{Bourgoin, M.}, \au{Baudet, C.}, \au{Cartellier,
  A.} \& \au{Gagne, Y.}} \yr{2007}  \at{Turbulent transport of material
  particles: An experimental study of finite size effects}.  \jt{Phys. Rev.
  Lett.}  \bvol{99}~(18).

\bibitem[Romano \& Antonia(2001)]{romano2001longitudinal}
{\sc \au{Romano, G.~P.} \& \au{Antonia, R.~A.}} \yr{2001}  \at{Longitudinal and
  transverse structure functions in a turbulent round jet: effect of initial
  conditions and {Reynolds} number}.  \jt{J.~Fluid Mech.}  \bvol{436},
  \pg{231--248}.

\bibitem[Sawford(1991)]{sawford1991reynolds}
{\sc \au{Sawford, B.~L.}} \yr{1991}  \at{Reynolds number effects in
  {Lagrangian} stochastic models of turbulent dispersion}.  \jt{Phys. Fluids~A}
   \bvol{3}~(6),  \pg{1577--1586}.

\bibitem[Sawford \& Yeung(2001)]{sawford2001lagrangian}
{\sc \au{Sawford, B.~L.} \& \au{Yeung, P.~K.}} \yr{2001}  \at{Lagrangian
  statistics in uniform shear flow: Direct numerical simulation and
  {Lagrangian} stochastic models}.  \jt{Phys. Fluids}  \bvol{13}~(9),
  \pg{2627--2634}.

\bibitem[Taylor(1922)]{taylor1922diffusion}
{\sc \au{Taylor, G.~I.}} \yr{1922}  \at{Diffusion by continuous movements}.
  \jt{Proc. Lond. Math. Soc.}  \bvol{20}~(1),  \pg{196--212}.

\bibitem[Toschi \& Bodenschatz(2009)]{toschi2009lagrangian}
{\sc \au{Toschi, F.} \& \au{Bodenschatz, E.}} \yr{2009}  \at{Lagrangian
  properties of particles in turbulence}.  \jt{Ann. Rev. Fluid Mech.}
  \bvol{41},  \pg{375--404}.

\bibitem[Vedula \& Yeung(1999)]{vedula1999similarity}
{\sc \au{Vedula, P.} \& \au{Yeung, P.~K.}} \yr{1999}  \at{Similarity scaling of
  acceleration and pressure statistics in numerical simulations of isotropic
  turbulence}.  \jt{Phys. Fluids}  \bvol{11}~(5),  \pg{1208--1220}.

\bibitem[Viggiano {\em et~al.\/}(2020)Viggiano, Friedrich, Volk, Bourgoin, Cal
  \& Chevillard]{viggiano2020modelling}
{\sc \au{Viggiano, B.}, \au{Friedrich, J.}, \au{Volk, R.}, \au{Bourgoin, M.},
  \au{Cal, R.~B.} \& \au{Chevillard, L.}} \yr{2020}  \at{Modelling {Lagrangian}
  velocity and acceleration in turbulent flows as infinitely differentiable
  stochastic processes}.  \jt{J.~Fluid Mech.}  \bvol{900},  \pg{A27}.

\bibitem[Volk {\em et~al.\/}(2011)Volk, Calzavarini, L\'ev\^eque \&
  Pinton]{volk2011dynamics}
{\sc \au{Volk, R.}, \au{Calzavarini, E.}, \au{L\'ev\^eque, E.} \& \au{Pinton,
  J.-F.}} \yr{2011}  \at{Dynamics of inertial particles in a turbulent von
  {K\'arm\'an} flow}.  \jt{J.~Fluid Mech.}  \bvol{668},  \pg{223--235}.

\bibitem[Volk {\em et~al.\/}(2008)Volk, Calzavarini, Verhille, Lohse, Mordant,
  Pinton \& Toschi]{volk2008acceleration}
{\sc \au{Volk, R.}, \au{Calzavarini, E.}, \au{Verhille, G.}, \au{Lohse, D.},
  \au{Mordant, N.}, \au{Pinton, J.-F.} \& \au{Toschi, F.}} \yr{2008}
  \at{Acceleration of heavy and light particles in turbulence: Comparison
  between experiments and direct numerical simulations}.  \jt{Physica~D}
  \bvol{237}~(14-17),  \pg{2084--2089}.

\bibitem[Voth {\em et~al.\/}(2002)Voth, {La Porta}, Crawford, Alexander \&
  Bodenschatz]{voth2002measurement}
{\sc \au{Voth, G.~A.}, \au{{La Porta}, A.}, \au{Crawford, A.~M.},
  \au{Alexander, J.} \& \au{Bodenschatz, E.}} \yr{2002}  \at{Measurement of
  particle accelerations in fully developed turbulence}.  \jt{J.~Fluid Mech.}
  \bvol{469},  \pg{121--160}.

\bibitem[Weisgraber \& Liepmann(1998)]{weisgraber1998turbulent}
{\sc \au{Weisgraber, T.~H.} \& \au{Liepmann, D.}} \yr{1998}  \at{Turbulent
  structure during transition to self-similarity in a round jet}.  \jt{Exp.
  Fluids}  \bvol{24}~(3),  \pg{210--224}.

\bibitem[Wolf {\em et~al.\/}(2012)Wolf, L\"uthi, Holzner, Krug, Kinzelbach \&
  Tsinober]{wolf2012investigations}
{\sc \au{Wolf, M.}, \au{L\"uthi, B.}, \au{Holzner, M.}, \au{Krug, D.},
  \au{Kinzelbach, W.} \& \au{Tsinober, A.}} \yr{2012}  \at{Investigations on
  the local entrainment velocity in a turbulent jet}.  \jt{Phys. Fluids}
  \bvol{24}~(10),  \pg{105110}.

\bibitem[Yeung(2002)]{yeung2002lagrangian}
{\sc \au{Yeung, P.~K.}} \yr{2002}  \at{Lagrangian investigations of
  turbulence}.  \jt{Ann. Rev. Fluid Mech.}  \bvol{34},  \pg{115--142}.

\bibitem[Yeung \& Pope(1989)]{yeung1989lagrangian}
{\sc \au{Yeung, P.~K.} \& \au{Pope, S.~B.}} \yr{1989}  \at{Lagrangian
  statistics from direct numerical simulations of isotropic turbulence}.
  \jt{J.~Fluid Mech.}  \bvol{207},  \pg{531--586}.

\bibitem[Zimmermann {\em et~al.\/}(2010)Zimmermann, Xu, Gasteuil, Bourgoin,
  Volk, Pinton, Bodenschatz \& {International Collaboration for Turbulence
  Research}]{zimmermann2010lagrangian}
{\sc \au{Zimmermann, R.}, \au{Xu, H.}, \au{Gasteuil, Y.}, \au{Bourgoin, M.},
  \au{Volk, R.}, \au{Pinton, J.-F.}, \au{Bodenschatz, E.} \& \au{{International
  Collaboration for Turbulence Research}}} \yr{2010}  \at{The {Lagrangian}
  exploration module: An apparatus for the study of statistically homogeneous
  and isotropic turbulence}.  \jt{Rev. Sci. Instrum.}  \bvol{81}~(5),
  \pg{055112}.

\end{thebibliography}

\end{document}